\documentclass[secnumarabic,amssymb, nobibnotes, aps, prd,onecolumn]{revtex4-2}
\usepackage{graphicx,epsfig,youngtab}

\pdfoutput=1
\usepackage{xcolor}
\usepackage{bbold}
\usepackage{amsmath}
\usepackage{hyperref}
\usepackage{amsfonts}
\usepackage{journal_names}
\expandafter\let\csname equation*\endcsname\relax 
\expandafter\let\csname endequation*\endcsname\relax 
\usepackage{bm}
\usepackage{lmodern}
\usepackage{moresize}
\usepackage{natbib}
\usepackage{tikz}
\usetikzlibrary{shapes.symbols,decorations.pathmorphing}
\usepackage{cabin}
\usepackage{pgfplots}
\pgfplotsset{compat = newest}

\def\q {\bm{q}}
\def\x {\bm{x}}
\def\r {\bm{r}}
\def\d {{\rm{d}}}

\def\x {\bm{x}}

\newcommand{\D} {{{\rm{D}}}}
\newcommand{\HH}{\mathcal{H}\,}

\renewcommand{\>}{\rangle}
\newcommand{\red}[1]{{\color{red}{#1}}}

\newcommand{\average}[1]{\left\langle #1 \right\rangle}

\newcommand{\doublesqbraket}[1]{\left[\left[ #1 \right]\right]}

\begin{document}

        % \title{\boldmath On the use of perturbation theory technique in the universe  on all scales } 
        % \title{\boldmath Beyond shell crossing with piecewise geodesics } 
          \title{\boldmath Vorticity generation in cosmology and the role of  shell crossing } 
        %\title{\boldmath Trapped surface for massive particles during  cluster  formation} 

\author{Obinna Umeh }
%\affiliation {Institute of Cosmology \& Gravitation, University of Portsmouth, Portsmouth PO1 3FX, United Kingdom}
\affiliation {${}^{1}$Institute of Cosmology \& Gravitation, University of Portsmouth, Portsmouth PO1 3FX, United Kingdom
%\\
%${}^{2}$Department of Physics, University of the Western Cape,Cape Town 7535, South Africa \\
%${}^{3}$Department of Mathematics and Applied Mathematics, University of Cape Town, Rondebosch 7701, South Africa}
%\email{
%obinna.umeh@port.ac.uk
}
%\email{obinna.umeh@port.ac.uk}
\email{umeobinna@gmail.com}
\date{\today}

\begin{abstract}

There is no source for cosmic vorticity within the cold dark matter cosmology.  However, vorticity has been observed in the universe, especially on the scales of clusters, filaments, galaxies, etc. 
Recent results from high-resolution general relativistic N-body simulation show that the vorticity power spectrum dominates over the power spectrum of the divergence of the peculiar velocity field on scales where the effective field theory of large-scale structure breaks down.  Incidentally, this scale also corresponds to the scale where shell-crossing occurs.   Several studies have suggested a link between shell crossing in the dark matter fluid and the vorticity generation in the universe, however, no clear proof of how it works {within general relativity} exists yet.  We describe for the first time how vorticity is generated in a universe such as ours with expanding and collapsing regions.  We show how vorticity is generated at the boundary of the expanding and collapsing regions. Our result indicates that the amplitude of the generated vorticity is determined by the {jump} in gradients of the gravitational potential, pressure and the expansion rate at the boundary.  
In addition, we argue that the presence of vorticity in the matter fields implies a non-vanishing magnetic part of the Weyl tensor. This has implications for the generation of Maxwell's magnetic field and the dynamics of clusters. The impact on accelerated expansion of the universe and the existence of causal limit for massive particles are discussed

\end{abstract}

\maketitle
\DeclareGraphicsRule{.wmf}{bmp}{jpg}{}{}
%\maketitle

\tableofcontents
\maketitle
%\newpage

\section{Introduction}

The standard cosmological model also known as the Lambda-Cold Dark Matter ($\Lambda$CDM) model has been successful in explaining some of the observed large-scale features of the Universe, for example, the observed anisotropies in the cosmic microwave background radiation~\cite{Smoot:1992td,Fixsen:1996nj}.  The model assumes that despite the inhomogeneous distribution of structures visible to an observer, the universe is well-approximated by the Friedmann-Lemaıtre–Robertson–Walker  (FLRW)  spacetime on  \emph{all length scales.}  It asserts that large-scale structures can be described as small perturbations on top of the background homogeneous,  isotropic  FLRW  spacetime. These perturbations can be decomposed into three modes: scalars, vectors and tensors. At the linear level, these modes propagate independently~\cite{Kodama:1985bj,Mukhanov:1990me}.  The scalar perturbations cannot induce a rotational part of the peculiar velocity. The vector perturbations can if there is a source initially but even if there is a source it decays very fast as the universe expands~\cite{Lu:2007cj}. The tensor perturbations can also induce the cosmic vorticity field but the amplitude is very small~\cite{jolicoeur:2018blf}.  At non-linear order, the evolution of the scalar perturbations can generate vector and tensor perturbations~\cite{Baumann:2007zm,Lu:2007cj}, however, their amplitude cannot explain the observed vorticity in large-scale structures~\cite{Barrera-Hinojosa:2020gnx,Barrera-Hinojosa:2021msx}. Vorticity can also be sourced by the entropy perturbations but the adiabatic perturbation appears to be preferred by the current observation \cite{Christopherson:2009bt}.

The cosmic vorticity field has been observed in galaxy clusters, filaments, galaxies, etc~\cite{Wang:2021axr}.  It is well known that most galaxies rotate and that the angular velocities of neighbouring galaxies are correlated~\cite{2019ApJ...872...78L}.
They play an important role in determining the observed galaxy spin and alignment~\cite{Krolewski:2019bfv}. On the solar system's scales, it dominates the dynamics of weather patterns~\cite{cushman2011introduction}, yet the evolution of the cosmic vorticity has no known source within the $\Lambda$CDM model of the universe. 
Within the general relativistic  N-body simulation, the vorticity power spectrum was recently estimated and it was found that it dominates over the power spectrum of the divergence of the peculiar velocity field of the matter field on scales where the effective field theory of large-scale structure breaks down~\cite{Jelic-Cizmek:2018gdp,Ivanov:2022mrd}. In this N-body simulation, the vector and tensor perturbations were turned off. Also, it was shown conclusively in the paper that the measured vorticity field is in the peculiar velocity of the matter as measured by the Eulerian observer~\cite{Adamek:2016zes}.  These details are important for the discussion that will follow. 

The work of \cite{Pueblas:2008uv} was the first to provide insights on how the generation of vorticity may be related to shell crossing of the matter fields.  Using the N-body simulation, they showed that the vorticity field tends to peak in the outskirts of virialised structures.  This result was confirmed by~\cite{Hahn:2014lca} using a different suit of N-body simulation. Hints of this were earlier discussed in \cite{Pichon:1999tk}, where the amount of vorticity generated after the first shell crossing in large-scale structure caustics was done.  There is no theoretical understanding of the connection between the shell-crossing of the mater fluid element and the generation of vorticity in the universe.  This lack of understanding motivated \cite{Cusin:2016zvu}  to consider whether the pair-wise velocity of galaxies can explain the observed vorticity. Similarly, the contribution of  higher cumulant of the phase-space distribution function was studied~\cite{Garny:2022tlk,Garny:2022kbk}. Their conclusion of \cite{Cusin:2016zvu} appears to show that the contribution of the pair-wise velocity is insufficient to explain the measured vorticity field. 

It is this gap that this paper plans to fill. We describe for the first time how cosmic vorticity field may be generated in the neighbourhood of "shell crossing singularity". We clarify the role of caustics or shell-crossing singularity in the generation of vorticity in cosmology.  To achieve this, we develop a model of the universe that describes more consistently the expanding and collapsing regions of universe. The standard cosmology model, describes only the expanding regions neglecting the gravitationally bound regions that have decoupled from the Hubble expansion.  We show that a consistent treatment of both regions is crucial to the generation of vorticity in the universe. 
The amplitude of the generated cosmic vorticity field depends on the difference between gradients of the gravitational potential, pressure  and convergence of flow lines between expanding and collapsing shells of matter.

The consequences of nonzero vorticity are enormous and we explored a couple of them. Firstly,  the existence of vorticity in the matter fields implies a non-vanishing magnetic part of the Weyl tensor associated with the matter flow velocity. The magnetic part of the Weyl tensor vanishes in Newtonian gravity, hence its measurement will constitute another test of general relativity. The nonzero magnetic part of the Weyl tensor could have implications for the existence of dark matter. This was studied in \cite{Ruggiero:2021lpf,Astesiano:2022ozl}. In addition, a non-zero cosmic vorticity field could also have implications for the generation and propagation of Maxwell's magnetic field in clusters. Finally, we show that differences in the expansion and contraction rates of the expanding and collapsing regions respectively could help explain the late-time accelerated expansion of the universe.

The rest of this paper is organised as follows: we describe particle trajectory,  fluid flow and shell crossing singularity in Newtonian gravity in section \ref{sec:geodesic_definition}.  We extended the same treatment to general relativity where we describe the geodesic of a massive particle on curved spacetime. We also identify the point where a geodesic ceases to be geodesic using the focusing theorem in this section. We show how the existence of an apparent horizon or the causal limit allows changing to a more appropriate coordinate since the Jacobian determinant at the causal limit is non-zero. The change of coordinate is possible because of the inverse function theorem. We introduce the model of the universe that describes the expanding and collapsing regions with appropriate boundary conditions consistently in section \ref{sec:universe_model}. We describe vorticity generation in section \ref{sec:vorticit_generation} and discuss other obvious implications of the model in section \ref{sec:implications} and conclude in section \ref{sec:conc}.

\section{Dynamics of massive particles on curved spacetimes}\label{sec:geodesic_definition}

In Newtonian gravity, the concept of massive particle motion is formulated in terms of the force the particles feel. In general relativity, massive particles travel along time-like geodesics.  Geodesics are not globally valid, especially in curved spacetime.  A curve could start out as geodesic but could cease to be geodesic in a finite time. Thus, the concept of geodesics on curved spacetime is only locally defined. 
 In this section, we shall describe in detail how shell crossing singularity or caustics in cosmology is related to the breakdown of the geodesic.  We also discuss how the formation of the causal horizon before caustic formation allows it to change to a more appropriate coordinate thereby avoiding the shell-crossing singularity.

\subsection{Fluid flow in Newtonian gravity}\label{sec:Newtonian_flow}

The modelling of the large-scale structures of the universe depends crucially on the solution of the geodesic equation.  The initial conditions for the Newtonian N-body simulations are set using the solution of the geodesic equation~\cite{Crocce:2005xz}. The force among particles distributed all over the universe is calculated using geodesic equation~\cite{Springel:2005mi}. One key point that is usually not mentioned when this approach is adopted is that a curve is geodesic only locally.  A geodesic can cease to be a geodesic within a finite time. This is usually not studied as a breakdown of geodesics in cosmology, however, it is a big area of research in Differential geometry~\cite{Sokolowski:2014non}.  Particles moving with non-relativistic velocities in the weak gravitational field regime in Newtonian gravity, its trajectory is given by
\begin{eqnarray}\label{eq:Newtonian_equation}
\frac{\d^2 x^i}{\d\tau^2}  &=& - {\nabla}^i \Phi({x^i})\,,
\end{eqnarray}
where ${\nabla}^i$ is a spatial derivative on Euclidean space, $\Phi$ is the gravitational potential, $\tau$ is proper time, $x^i$ is the position of the particle or fluid element at $\tau$.  $x^i$ is related to the initial position, $q^i$, of the particle according to 
\begin{eqnarray}
x^i(\tau,q^i) = q^i + \Psi^i(\tau,q^i)\,,
\end{eqnarray}
where $\Psi^i$ is known as the displacement field. 
It is customary to describe $x^i$ and $q^i$ as the Eulerian and Lagrangian coordinates respectively.
Initially,  that is at $\tau = 0$, $\Psi^i(q,\tau)=0$. In the expanding universe,  it is beneficial to work in comoving coordinates  $  r^i = x^i/a(\eta)$, where $a$ is the scale factor of the universe,  $\eta$ is the conformal time, it is related to the proper time according to $ \d \eta =\d \tau /a(\eta)$. The gravitational potential is related to the mass density, $\rho$ through the Poisson equation $\nabla^2 \Phi = 4\pi a^2 \delta \rho_{m}$, where $\delta \rho$ is a perturbation in the mass-density with respect to a background FLRW  value $\bar{\rho}$.  Imposing conservation of mass-density, that is  $\bar{\rho}(\tau) \d^3 q = \rho(\tau, {x^i})\d^3 x$, leads to 
\begin{eqnarray}\label{eq:mass_conservation}
1+ \delta(x,\tau) = \bigg|\frac{\d^3 x}{\d^3 q}\bigg| ={ \frac{1}{J(\tau,\q)}}\,,
\end{eqnarray}
where $\delta \equiv \delta \rho/ \bar{\rho}$ is the density contrast or fluctuation in the mass-density and $J = {\rm{det}}  \left[ \delta_{ij} + \Psi_{i,j}\right]$ is the Jacobian of the transformation.  According to Zel'dovich \cite{Zel'dovich:1970AandA},  the leading order approximation to $J$ is given by
\begin{eqnarray}\label{eq:Jacobian_matrix}
J(\tau,{q}) = \begin{vmatrix}
 1-\alpha D_1(\tau) &0&0 \\
0&1-\beta D_1(\tau)  &0  \\
 0 & 0&  1-\gamma D_1(\tau) 
\end{vmatrix}\,,
\end{eqnarray}
where $\alpha$, $\beta$ and $\gamma$ are eigenvalues and $D_1$ is the matter density growth function.  The caustics occurs, i.e $J\to 0$ whenever   $ 1-\alpha D_1(\tau) \to0$.   According to equation \eqref{eq:mass_conservation}, the density contrast diverges at a caustics  $1+ \delta(x,\eta)\to \infty$. 
This is sometimes interpreted as an indication of a breakdown of Zel'dovich approximation.  However, going beyond the Zeldovich approximation by adding higher-order perturbation theory terms does not resolve or remove the caustics~\cite{Garrison:2016vvp}. 
In fact, it is more of an indication of a breakdown of the 'one-parameter' cosmological perturbation theory. It is important to mention that it is more of an indication of a breakdown of one-parameter cosmological perturbation theory because cosmological perturbation theory can be formulated in many-parameters by dynamically switching to a most suitable background spacetime~\cite{Bruni:2002sma,Nakamura:2003wk,Goldberg:2016lcq}. 
In the language of fluid dynamics, it is an indication of a breakdown of the single-stream approximation~\cite{Rampf:2020hqh}. It is also clear in this language that when the single-steam approximation breaks down, the natural progression towards progress is to introduce the two-stream approximation~\cite{1980JAtS...37..630M,1987AmJPh..55..524B}. 
%In Cosmology, apart from the efforts of \cite{Goldberg:2016lcq}

There have been several attempts towards developing a consistent two-stream approximation or two-parameters perturbation theory in cosmology~\cite{Bruni:2002sma,Nakamura:2003wk,Goldberg:2016lcq,Gallagher:2019tcr,Clifton:2020oqx}. These attempts focus on the impact of the coupling between small and large-scale dynamics on the large-scale features of the universe. The results so far appear to show that the coupling could be important for the gravitational waves~\cite{Gallagher:2018bdl}. Consistent treatment of the boundary condition for the scalar perturbations which dominate dynamics on small scales is missing. One surprising attribute of these approaches is that they assume a priori that Newtonian approximation is valid on small scales~\cite{Goldberg:2017gsm,Clifton:2020oqx}. This is one of the crucial points that we highlight here. Newtonian gravity does not appear to describe some of the critical events preceding the formation of caustics. For example, it has been pointed out earlier that a massive particle comoving with the Hubble expansion cannot influence dynamics within a virialised local environment if there exists a causal horizon~\cite{Ellis:2010fr}.  Newtonian gravity lacks the tools to describe the formation of causal horizons~\cite{HawkingandEllis:1973lsss.book}. There have been works attempting to justify the use of Newtonian gravity in cosmology on all scales~\cite{Hwang:2005wb}.  These claims are yet to account for some degrees of freedom in general relativity such as the magnetic part of the Weyl tensor that is fundamentally not contained in Newtonian gravity \cite{Ehlers:2009uv,Clifton:2016mxx}.  The magnetic part of the Weyl tensor is non-zero if the vorticity is non-zero~\cite{Ellis:1998ct}. 

As a result of these, the rest of our discussion will be based on general relativity. It is possible to realise some of these features that precede the formation of caustics in Newtonian gravity after the fact. 
%We describe some of these in  Appendix \ref{sec:focusing_theorem_newtonian_limit}.  
Furthermore, general relativity is consistent with the principle of least action. The principle of least action is central to our model building because it is straightforward to derive consistent boundary conditions for geodesics using the principle of least action.  Some of the ideas we discuss here are similar to those employed in the study of gravitational memory effect but our approach is more fundamental~\cite{Kumar:2021qrg}.

\subsection{ Geodesic of relativistic massive  particle and its evolution equation }\label{sec:Geodesic_equation}

The action of a relativistic massive particle is minus the rest energy times the change in time $S= - E\Delta \tau =- m \Delta \tau$, where $\Delta \tau = \tau_{f} - \tau_{i}$, where $\tau_{i}$ is the initial time when the seed was created and $\tau_{f}$ is the future time. The particle follows a maximal geodesic as it evolves from $\tau_{\rm{ini}}$ to $\tau_f$. The most essential point here is that the concept of a maximal geodesic is only locally defined in curved spacetime.  %In Newtonian gravity, there is really no distance, geodesics is considered a global property.  
Therefore, it is essential to ascertain the range of validity of geodesic. 

Let $\gamma$ denote a smooth time-like  curve defined within an interval $[\tau_{\rm{ini}},\tau_f]$ on Pseudo-Riemann manifold $\mathcal{M}_4$ in 4 dimensions. The massive particle action is given by 
\begin{eqnarray}\label{eq:length_functional}
S(\gamma)  = -m \int_{\tau_i}^{\tau_f} L(\gamma(\tau), \dot{\gamma}(\tau)) \d\tau  = -m\int_{\tau_i}^{\tau_f}\sqrt{-g_{ab} \frac{ \d x^a }{\d\tau } \frac{\d  x^b }{\d \tau }}   \d\tau \,, ~~~~
{\rm{with}}~~~~ L = \sqrt{-g_{ab} \frac{\d x^a }{\d\tau}\frac{\d x^b}{\d\tau} }\,,
\end{eqnarray}
where $g_{ab}$ is the metric of the spacetime on $\mathcal{M}_4$, $L(\dot{\gamma}(\tau), \dot{\gamma}(\tau)) $ is the Lagrangian and $\dot{\gamma} = \d \gamma/\d\tau$. In the second equality, we introduced  $x^a $, i.e the coordinates of the points on the manifold. We will drop the \red{$-m$} in equation \eqref{eq:length_functional} for the rest of the presentation to reduce clutter. 
Equation \eqref{eq:length_functional} is invariant under reparametrisation. For  $\gamma$ to be a geodesic within $[\tau_{\rm{ini}},\tau_f]$, it has to be the critical point of the infinitesimal variation of $\gamma(\tau)$. 
Let  $x^a(\tau,s) = \bar{x}(\tau) = s\delta x (\tau)$ be a variation, where $s$ parameterises nearby curves  $s \in (-\epsilon, \epsilon)$. The central geodesic is given by $\bar{x}^a(\tau) = x^a(\tau,0)$.
 A variation is proper when all the nearby curves converge at the end-points
\begin{eqnarray}\label{eq:proper_variation}
x^a(\tau_{\rm{ini}},s) =\bar{x}^a(\tau_{\rm{ini}}), \qquad x^s(\tau_f,s) = \bar{x}(\tau_f) \,.
\end{eqnarray}
On curved spacetime, a family of nearby curves could converge before the endpoint. When this happens a geodesic will no longer maximax the action if it is extended beyond this point. The point where a family of geodesics converge before the end-point is called a conjugate point. A conjugate point is a caustic since the Jacobian vanishes there. 
Let's introduce a time-like  4-velocity, $u^a$ and a deviation vector, $\xi^a$ that tracks the propagation of the nearby family of curves:
\begin{eqnarray}\label{eq:geodesic_equation}
u^a = \frac{\d x^a(\tau,s)}{\d\tau}, ~~~~~ \qquad \xi^a = \frac{\partial x^a(\tau,s)}{\partial s}\,.
\end{eqnarray}
It is well known that the first variation of an action is equivalent to taking the functional derivative of the action. In our case, the action in \eqref{eq:length_functional} has some amazing symmetry by the Noether theorem, i.e the Lagrangian does not depend explicitly on $x^a$. This implies that $\xi^a$ can be Lie dragged along $u^a$: $ u^b\nabla_{b} \xi^a =\xi^b \nabla_{b} u^a  $
\begin{eqnarray}
 \frac{\d S}{\d s}\bigg|_{s=0} & =&\int_{\tau_{\rm{ini}}}^{\tau_i} \frac{\partial }{\partial s} \left( L(\dot{x}^a, \dot{x}^a)\right)\d\tau=-\int_{\tau_{\rm{ini}}}^{\tau_i} \frac{u_a}{L} \xi^c\nabla_{c} u^a \d\tau =-\int_{\tau_{\rm{ini}}}^{\tau_i} \frac{u_a}{L} u^c\nabla_{c} \xi^a \d\tau 
 \label{eq:firstline_action}
 \\
 & =&\int_{\tau_{\rm{ini}}}^{\tau_i} \frac{\d}{\d\tau} \left[ \xi_{a} \frac{u^a}{L}\right] \d\tau+ \int_{\tau_{\rm{ini}}}^{\tau_i}  {\xi_a} u^b\nabla_{b}\left[ \frac{ u^a}{L} \right]\d\tau
\,,
\\
 &=&\int_{\tau_{\rm{ini}}}^{\tau_i}  \xi_c u^d\nabla_d \left[ \frac{u^c}{L}\right]  \d \tau
 + \left(\frac{ \xi_c u^c}{L} \right)\bigg|_{\tau_{\rm{ini}}}^{\tau_{f}}
 = \int_{\tau_{\rm{ini}}}^{\tau_i}  \left[\frac{\xi_c }{L} \right]u^d\nabla_d u^c \d \tau\,,
 \label{eq:variation}
\end{eqnarray}
where we have imposed the orthogonality condition $\xi_a u^a = 0$ in the last line.  And in the third line, we imposed the conditions for proper variation of the action. One can check that the result in equation \eqref{eq:variation} is correct by simply taking the functional derivative of an action with an arbitrary Lagrangian which gives
\begin{eqnarray}\label{eq:functional}
 \frac{\d S}{\d s}\bigg|_{s=0} &=& \ \frac{\partial. L}{\partial \dot{x}^a}\bigg|_{\tau_{\rm{ini}}}^{\tau_{f}} +   \int_{\tau_{\rm{ini}}}^{\tau_i}\left[ \frac{\partial L}{\partial x^a}(\tau) - \frac{\d}{\d \tau} \frac{\partial L}{\partial {\dot{x}^a} }    \right]\xi^a(\tau)\d \tau \,.
\end{eqnarray}
Putting the Lagrangian in equation \eqref{eq:length_functional} into equation \eqref{eq:functional} gives the same result. 
The critical point of  these equations  (i.e $ {\d S}/{\d s}|_{s=0} =0$ gives the geodesic equation  $u^c \nabla_{c} u^a = 0$. 

\subsection{ Validity of geodesics, focusing theorem and horizon }\label{eq:GDE_variation}

To understand when a geodesic ceases to be a geodesic, we need a second derivative test. This is also called the second variation of the action. 
The second variation measures how fast nearby geodesics are expanding or contracting towards the central geodesics $\gamma_{0}(\tau)$. The second variation of equation \eqref{eq:length_functional} may be obtained by simply taking the derivative of equation \eqref{eq:firstline_action}
\begin{eqnarray}
 \frac{\d^2 S}{\d s^2}\bigg|_{s=0} &=&  -\int_{\tau_{\rm{ini}}}^{\tau_i} \frac{\partial }{\partial s} \left[ \frac{u_a}{L} u^c\nabla_{c} \xi^a \right]\d\tau 
 \\
 &=& -\int_{\tau_{\rm{ini}}}^{\tau_i} \left[ \frac{u_a}{L} \right] \nabla_{\xi} \nabla_{u} \xi^a \d\tau  -\int_{\tau_{\rm{ini}}}^{\tau_i} \nabla_{\xi}\left[ \frac{u_a}{L} \right]  \nabla_{u} \xi^a \d\tau \,,
\end{eqnarray}
where we have introduce a shorthand notation for directional derivatives $\xi^c\nabla_{c} = \nabla_{\xi}$ and $u^c\nabla_{c} = \nabla_{u}$. The index position on the first term can be switched with the help of the Ricci identity
\begin{equation}
2\nabla_{[\xi}\nabla_{u]} \xi_a  = R^d{}_{abc}  u^a \xi^ c{\xi}_{d} \rightarrow  \nabla_{\xi}\nabla_{u} \xi_a = \nabla_{u}\nabla_{\xi} \xi_a +  R^d{}_{abc}  u^a \xi^ c{\xi}_{d}\,.
\end{equation}
After some simplification, it gives
\begin{eqnarray}
 \frac{\d^2 S}{\d s^2}\bigg|_{s=0} &=& -\int_{\tau_{\rm{ini}}}^{\tau_i} \left[ \frac{u_a}{L} \right] \left[ R^d{}_{abc}  u^a \xi^ c{\xi}_{d}+ \nabla_{u}\nabla_{\xi} \xi_a  \right] \d\tau  -\int_{\tau_{\rm{ini}}}^{\tau_i} \nabla_{\xi}\left[ \frac{u_a}{L} \right]  \nabla_{u} \xi^a \d\tau \,.
\end{eqnarray}
Performing integration parts leads to 
\begin{eqnarray}\label{eq:second_variation}
- \frac{\d^2 S}{\d s^2}\bigg|_{s=0} &=&\int_{\tau_{\rm{ini}}}^{\tau_i}  \xi_{c} \left[ \frac{\d^2 \xi^c}{\d\tau^2} + R^{c}{}_{def} \xi^{d} u^{e} u^{f} \right]\d\tau
+\left[ u_c \xi^b\nabla_{b}\xi^c\right]\bigg|_{\tau_{\rm{ini}}}^{\tau_{f}} +\left[ \frac{\d \xi ^a}{\d\tau } \xi_a\right] \bigg|_{\tau_{\rm{ini}}}^{\tau_{f}}\,,
\end{eqnarray}
%\frac{\partial f^c}{\partial s}\right]
where $ R^{a}{}_{def}$ is the Riemann tensor and ${\D \xi ^a}/{\D\tau }  = u^a \nabla_a \xi^b$. The equation of motion resulting from the first variation has been imposed.  Further imposing the condition for proper variation $\xi^a(p) = \xi^a(q) = 0$, equation \eqref{eq:second_variation} reduces to 
\begin{eqnarray}\label{eq:second_variation2}
- \frac{\d^2 S}{\d s^2} &=&\int_{\tau_{\rm{ini}}}^{\tau_i}  \xi_{c} \left[ \frac{\d^2 \xi^{c}}{\d\tau^2} + R^{c}{}_{def} \xi^{d} u^{e} u^{f} \right]\d\tau \,.% \quad\to \quad \frac{\D^2 \xi^{c}}{\D\tau^2} + R^{c}{}_{def} \xi^{d} u^{e} u^{f}  = 0\,.
\end{eqnarray}
%\subsection{Breakdown of single stream regime on  curved spacetimes}
The critical point (i.e $ {\d^2 S}/{\d s^2} = 0)$ gives the geodesic deviation equation. 
\begin{eqnarray}\label{eq:geodesic_deviation}
 \frac{\d^2 \xi^{c}}{\d\tau^2} + R^{c}{}_{def} \xi^{d} u^{e} u^{f}  = 0 \,.
\end{eqnarray} 
We can also obtain the same geodesic deviation equation in equation \eqref{eq:geodesic_deviation}  using the following Lagrangian in equation \eqref{eq:functional} and setting $\xi^b\nabla_{b}\xi^c =0$:
\begin{eqnarray}\label{eq:Lagrangian_GDE}
L(\xi^a,\dot{\xi}^a) =  \eta_{ab} \frac{\d \xi^a}{\d \tau}  \frac{\d \xi^b}{\d \tau}  - \frac{1}{2} R_{abcd} u^a u^b \xi^b \xi^d \,.
\end{eqnarray}
This connection will be important later. 
It is easier to extract information from equation \eqref{eq:geodesic_deviation} by decomposing the spacetime into temporal and the spatial part, the most consistent way to do this is to consider foliations where $\xi^a$ Lie dragged along the integral curves of $u^a$~\cite{HawkingandEllis:1973lsss.book}.
    \begin{eqnarray}\label{eq:firstorderode}
\mathcal{L}_{u} \xi^a = 0 \rightarrow  \frac{{\d} \xi^{a}}{{\d} \tau}  = \nabla_{b} u^a \xi^b  =B_{b}{}^{a}\xi^b\,,
  %  = \left[\frac{1}{3}\Theta h^a{}_{b }+\sigma^{a}{}_{b} + \omega^{a}{}_{b}\right] Z^b \,,%= \chi_{b a}  Z^b
  \end{eqnarray}
where $B_{ab} = \nabla_{b} u_{a}$. $B_{ab}$ measures the deformation of the curved spacetime in comparison to flat space.  It can be decomposed into irreducible coordinate independent components~\cite{Ellis:1998ct,Ellis:1990gi}
\begin{eqnarray} \label{eq:covdu}
B_{ab} = \nabla_{b}u_{a}=-u_{b}A_{a}+\frac{1}{3}\Theta h_{ab }
+\sigma_{ab} + \omega_{ab}\,,
\end{eqnarray} 
 where  $A_a$ is the acceleration, $\Theta$  describes the expansion of the one-parameter family of geodesics if $\Theta >0$ and contraction or collapsing of one-parameter family of geodesics if $\Theta<0$.  $\sigma_{ab}$ called the shear tensor, describes the rate of change of the deformation of a one-parameter family of geodesics when compared to flat spacetime.  $\omega_{ab}$ is the vorticity tensor. it is an anti-symmetric tensor, It describes the twisting of a one-parameter family of nearby geodesics.   We define also the  scalar invariant of these tensors as follows, for the shear tensor ${\sigma}^2 = {\sigma}_{ab} {\sigma}^{ab}/2$ and the vorticity tensor
${\omega}^2 = {\omega}_c{\omega}^c/2={\omega}_{ab}{\omega}^{ab}/2)$, where $\omega_{a} = \frac{1}{2}\varepsilon_{abc} \omega^{bc}$ is a vorticity vector. $h_{ab}$ is the metric on the hypersurface. It is defined in terms of the metric $g_{ab}$ and $u^a$
\begin{eqnarray} 
{h}_{ab}={g}_{ab}+{u}_a {u}_b,~~~~~~~~ ~ {\epsilon}_{ab c}={\eta}_{ab c d}u^d\,,
\end{eqnarray}
where ${\eta}_{abcd} = 2{u}_{[a}{\epsilon}_{b]cd}-2{\epsilon}_{ab[c}{u}_{d]}$ is the alternating tensor for the full spacetime\cite{Ellis:1998ct}. 
Note that  $B_{ab}$ is related to the extrinsic curvature tensor according to $K_{ab} = h_{a}{}^{c} h_{b}{}^{d}B_{cd}$.
% where $\varepsilon_{\pm} =\pm1$..
Putting equation \eqref{eq:firstorderode} into equation \eqref{eq:geodesic_deviation}, we find that these geometric quantities satisfy the following equations~\cite{2012reco.book.....E,Ellis1971grc..conf..104E,Ellis2009,Ellis:1998ct}:
\begin{eqnarray}
\frac{ {\rm{D}} {{\Theta}} }{{\rm{D}} \tau} &=& - \frac{1}{3}{\Theta}^2 - {\sigma}_{ab}{\sigma}^{ab} -{\omega}_{ab}{\omega}^{ab} +{\D}_{a} A^a + A_a A^a - {R}_{ab} {u}^a {u}^b\,,
\label{eq:expansion}
\\
\frac{ {\rm{D} } {{\sigma}}_{ab}}{{\rm{D}} \tau} &=& - \frac{2}{3} {\Theta} {\sigma}_{ab} - {\sigma}^{c}{}_{\<a}{\sigma}_{b\>c} - {\omega}^{c}{}_{\<a}{\omega}_{b\>c}+{\D}_{\<a} A_{b\>}  +A_{\<a}A_{b\>} - {C}_{acbd}{u}^c {u}^d\,,
\label{eq:shear}
\\
\frac{ {\rm{D} } {{\omega}}_{ab}}{{\rm{D}} \tau} &=& -\frac{2}{3}{\Theta}{\omega}_{ab} + {\sigma}^c{}_{[a}{\omega}_{b]c} - {\D}_{[a} A_{b]}
\label{eq:vorticity}\,,
\end{eqnarray} 
where ${C}_{acbd}$ is the Wely tensor  and $R_{ab}$ is the Ricci tensor.  Equations \eqref{eq:expansion},  \eqref{eq:shear} and \eqref{eq:vorticity} can also be derived using the Ricci identity.  General relativity is needed to relate the  Ricci tensor in equation \eqref{eq:expansion} to the energy-momentum tensor. We make minimal assumptions about the form of the energy-momentum tensor.
The Weyl curvature tensor may be decomposed further into the electric ${E}_{ab}$ and the magnetic part ${H}_{ab}$  with respect to $u^a$
\begin{eqnarray}\label{eq:Weyl}
{C}_{ab}{}{}^{cd}&=&		
4\left({u}_{[a}{u}^{[c}+{h}_{[a}{}^{[c}\right){E}_{b]}{}^{d]}
%\\ \nonumber&&
+2{\varepsilon}_{abe}{u}^{[c} {H}^{d]e}+ {u}_{[a}{H}_{b]e}{\varepsilon}^{cde}\,,
\end{eqnarray}
where  ${E}_{ab}$ and  ${H}_{ab}$  are defined as ${E}_{ab} =  {C}_{acbd} {u}^c{u}^d $ and ${H}_{ab}= \,{\varepsilon}_a{}^{cd}{C}_{cdbe}{u}^e/2$.  ${E}_{ab}$ and  ${H}_{ab}$ live on the hypersurface $E_{ab} u^{b} = 0=H_{ab} u^{b} $.
Note that ${E}_{ab}$ describes the impact of the tidal forces due to local massive distribution, while the magnetic part describes the tidal forces due to the twisting or stretching of spacetime along different directions.
In the Newtonian limit, $E_{ij} \approx \partial_{i} \partial_{j} \Phi$, where   $\Phi$ is the gravitational potential and $E_{ij} = \partial_{\<i} \partial_{j\>} \Phi =  \partial_{(i} \partial_{j)} \Phi -\partial^2 \Phi/3$ and ${H}_{ab}$ vanishes in the Newtonian limit. 
The most well-known consequence of the above equation is the prediction that the vorticity vanishes exactly.  This can be seen by expressing the derivative in equation \eqref{eq:vorticity} in terms of the Lie derivative
\begin{eqnarray}\label{eq:Lieomea}
\mathcal{L}_{u} \omega_{ab} = \frac{\text{D} {{\omega}}_{ab}}{\text{D} \tau}     +\frac{2}{3}{\Theta}{\omega}_{ab} - {\sigma}^c{}_{[a}{\omega}_{b]c} ={\D}_{[a} A_{b]}\,.
\end{eqnarray}
In the gravitational rest frame $A^a$ is a gradient of a scalar $A _a = \nabla_{a} \Phi$, hence ${\D}_{[a} A_{b]} = 0$.
Therefore,  irrespective of the coordinate system, $\omega_{ab}$ vanishes if the initial vorticity is zero. 
%This implies that the vorticity vanishes if it was zero initially. 
The vanishing of the vorticity or the existence of the vorticity-free congruence implies $u^a$ is hypersurface orthogonal. 
 This also means that $u^a$ maybe derived from the covariant of a scalar field $S$: $
u_{a} = - {\nabla_{a} S}/{||\nabla S||}$,
where $||\nabla S||$ is a normalization factor and $u^a$ is pointing in the future direction.

The shear propagation equation may be written in a coordinate-independent form  as 
\begin{eqnarray}
\mathcal{L}_{u} \sigma_{\mu \nu} &=& \frac{{\text{D}} \sigma_{\mu\nu}}{\text{D}\tau} +\frac{2}{3} \sigma_{ab} \Theta 
+2 \sigma_{\<a}{}^{c} \sigma_{b\>c} +\omega_{\<a}{}^{c} \sigma_{b\>c}
={\D}_{\<a} A_{b\>}  +A_{\<a}A_{b\>} +\sigma_{\<a}{}^{c} \sigma_{b\>c}- E_{ab}
+ \frac{1}{2} R_{\<ab\>}\,.
\end{eqnarray}
The shear is sourced by the electric part of the Weyl tensor and the trace-free part of the Ricci tensor.
\begin{eqnarray}\label{eq:shearprop2}
 \mathcal{L}_{u} \sigma_{\mu \nu}   =  \tilde{\sigma}^{c}{}_{\<a}\tilde{\sigma}_{b\>c}  - E_{ab}
+ \frac{1}{2} R_{\<ab\>} \,.
\end{eqnarray}
Finally, from equation \eqref{eq:expansion} we can obtain the time-like geodesic version of the focussing theorem\red{~\cite{HawkingandEllis:1973lsss.book,Poisson:2003nc}}%\cite{Poisson:2003nc}.
 From equation \eqref{eq:shearprop2}, it is clear that $\sigma_{ab}\sigma^{ab} \ge  0, $ for zero vorticity and assuming that the weak  energy condition holds that is  $ {R}_{ab} {u}^a {u}^b \ge 0$  \cite{Senovilla:2006db}, then  equation \eqref{eq:expansion} becomes 
\begin{eqnarray}\label{eq:focussing_theorem}
\frac{ {\rm{D}}  \Theta}{ {\rm{D}} \tau}  \leq -\frac{\Theta^{2}}{3} \,.%\rightarrow  \frac{1}{\Theta } \geq \frac{1}{\Theta _{0}}+ \frac{\tau }{3}
\end{eqnarray}
Integrating  with respect to $\tau$  gives ${ { {1}/{\Theta }}\geq {{1}/{\Theta _{0}}}+{ {\tau }/{3}}}$,
where  ${ \Theta _{0}}$ is the initial value of the expansion.  Equation \eqref{eq:focussing_theorem} describes the features of geodesics that must collapse to caustics. Such geodesics must be collapsing initially. We go into greater detail in sub-section \ref{subsec:separate_universe} to describe how this happens in cosmology.

\subsection{Inevitability of  more than one-parameter family of curves in a universe like ours }\label{subsec:separate_universe}

Within the standard model of cosmology,  the cosmological inflation models are usually built on an FLRW background spacetime. The model predicts the initial conditions for the large-scale structures of the universe. The current observation suggests that the seeds of large-scale structure formation follow a Gaussian distribution~\cite{Aghanim:2018eyx}. General relativity is a deterministic theory, hence,  it is possible to probe models of cosmological inflation using the observations in the late universe~\cite{Koyama:2018ttg,Maartens:2020jzf}.  One could extend this argument to the focusing theorem(equation \eqref{eq:focussing_theorem}), that the family of geodesics that collapses to form clusters, galaxies we see today are those that found themselves within over-dense regions at the initial time, while those that found themselves in under-dense regions evolve to form voids. Evolution histories of these regions are different as we will show in section \ref{sec:universe_model}. The concept of tracer bias in modelling the clustering of large-scale structures is based on this idea~\cite{Verde:2009hy,Dai:2015jaa}.
The discussion below goes into greater detail to describe the distinction between the one-parameter family of geodesics that collapses to clusters and the one-parameter family of geodesics that evolves to form voids. 

Equation \eqref{eq:firstorderode} is a linear first-order differential equation, hence, the solution  at present time 
 is related to  its values at some point $q$ in the past according to 
$ \xi^{i}(\tau,{\x}) = J_{ij}(\tau,{\x}) \xi^{j}(\tau_{\rm{ini}},{\q})\,,$
 where $ J_{ij}$ is a Jacobi  matrix and $i,j$ indicates components.  
 %Just like in the Newtonian limit, the Jacobian $J = {\rm{det}}[J_{ij}]$ determines shell crossing. In GR, it contains more information. 
 Putting $ \xi^{i}({\x}) = J_{ij}({\x}) \xi^{j}({\q})\,,$ in equation  \eqref{eq:firstorderode} gives 
 \begin{eqnarray}\label{eq:J-S-relationship}
 \frac{\D J_{i}{}^{j}({\x}) }{\D\tau} = K_{i k } J^{jk}\,,
 \end{eqnarray}
 where, $K_{ab} =h_{b}{}^{c} \nabla_{c} u_{a}$ is the extrinsic curvature of the hypersurface orthogonal to $u^a$,
  Although the relationships between various components of $ J_{i}{}^{j}$ and $K_{i k } $ are important(see \cite{Clarkson:2011br} in the case null geodesics),  our interest at the moment  is on the determinant which is given by
\begin{eqnarray}\label{eq:volumeelement}
\frac{1}{{{\rm{det}}[J](\tau, {\x})} }\frac{{\rm{D}} {\rm{det}}[J](\tau, {\x})}{{\rm{D}} \tau} &=&  \Theta(\tau, {\x}) \,,
%=\frac{1}{\sqrt{{\rm{det}}[h_{ab}](\tau, {\x})} }\frac{{\rm{d}} \sqrt{{\rm{det}}[h_{ab}](\tau, {\x})}}{{\rm{d}} \tau}\,.
\end{eqnarray}
where ${\rm{det}[J]}$ is the determinant of the Jacobian.  {In order to obtain this equation, we made use of the Jacobi formula~\cite{bellman97}, which  expresses the derivative of the determinant of any matrix ${\bf{A}}$ whose inverse exists  in terms of the adjugate of ${\bf{A}}$ and the derivative of ${\bf{A}}$. In relation to equation \eqref{eq:J-S-relationship}, we are assuming that there are no caustics ${\rm{det}}[J](\tau, {\x}) \neq 0$ .}
  Integrating equation \eqref{eq:volumeelement} gives
  \begin{eqnarray}\label{eq:censorrship}
{\rm{det}}[J](\tau,{\x}) = {\rm{det}}[J](\tau_{\rm{ini}},{\q}) \exp\left[ \int_{\tau_{\rm{ini}}}^{\tau} \d\eta' \Theta(\tau' ,{\x}({\tau})) \right]\,.
\end{eqnarray}
 At  $ \Theta= 0$,  the Jacobian becomes ${\rm{det}}[J](\tau, {\x}_{\rm{dp}}) = {\rm{det}}[J](\tau_{\rm{ini}}, {\x}_{\rm{dp}})$, where ${\x}_{\rm{dp}}$ defines  a spatial location where  $ \Theta = 0 $.  At this location, the property of the fluid element changes; it is incompressible fluid at ${\x}_{\rm{dp}}$~\cite{Fodor:1998jp,cushman2011introduction}.
Furthermore, equation \eqref{eq:volumeelement} has the form of an autonomous differential equation (${\d y}/{\d \tau } = F(\tau ,y)$), hence one could argue that $\Theta= 0$ is a critical point. In the observed universe, there are locations where $\Theta(\tau, {\x}_{\rm{dp}})= 0$. This is usually measurable in peculiar velocity surveys~\cite{Karachentsev:2008st,Anand:2019ApJ...880...52A,Qin:2021tak}. For the local group observer, this location is known as the zero-velocity surface and its radius has been determined precisely~\cite{2006Ap.....49....3K,Karachentsev:2008st,1999AandARv...9..273V}. The consequences of this for the supernova absolute magnitude tension were explored in~\cite{Umeh:2022kqs,Umeh:2022prn,Umeh:2022kqs}.
Our discussion here is more general, we treat these as critical location or causal horizon that indicates an end to a collapsing region of the universe and the beginning of an expanding region for an observer in a virialised region such as ours. The causal horizon can easily be  determined by splitting $\Theta$ into  two parts: 
\begin{eqnarray}
\Theta = \Theta_{H}+ \Theta_{L} \,,
\end{eqnarray}
where $\Theta_H$ denotes the expanding part, i.e $\Theta_H = 3 H$(this is determined by the background FLRW spacetime) and  $\Theta_{L}$  describes the local component. 
The expanding component is always positive, $\Theta_{H}>0$, while the local component could be positive, negative and zero.  $\Theta_{L}=0$ implies a universe without large-scale structures or the large-scale structures have zero peculiar velocity relative to Hubble expansion, $\Theta_{L}>0$, implies an equally expanding local regions. The relative dominance of $\Theta_{H}$ and $\Theta_{L}$ divides the universe into expanding and collapsing regions.
The locations in the universe where the gravitational field of gravitationally bound structures dominate  $\Theta_{L} >\Theta_H $(e.g. haloes, etc) define the collapsing regions. The star and galaxy formation happen within this region~\cite{Ellis:2010fr}.
Within this region, the one-parameter family of geodesics are converging to a singularity/caustic according to the focusing theorem, however, nonlocal effects may intervene to prevent a singularity formation~\cite{HawkingandEllis:1973lsss.book}. The locations where $\Theta_{L} <\Theta_H$ (e.g. void, vacuum) are expanding and will continue to expand.  These are the expanding regions.

\begin{figure}[h]
\centering 
\includegraphics[width=80mm,height=70mm] {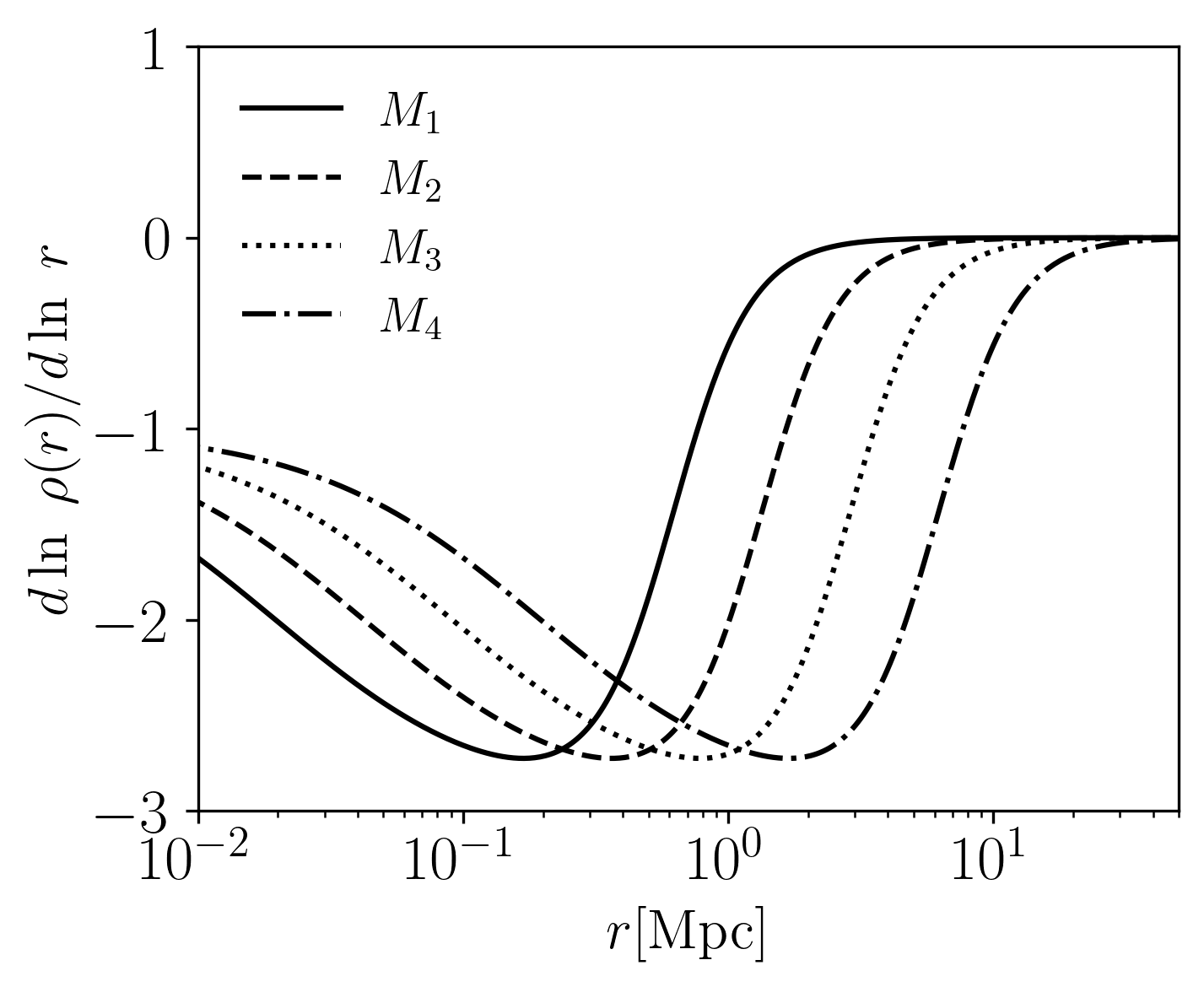}
\includegraphics[width=80mm,height=70mm] {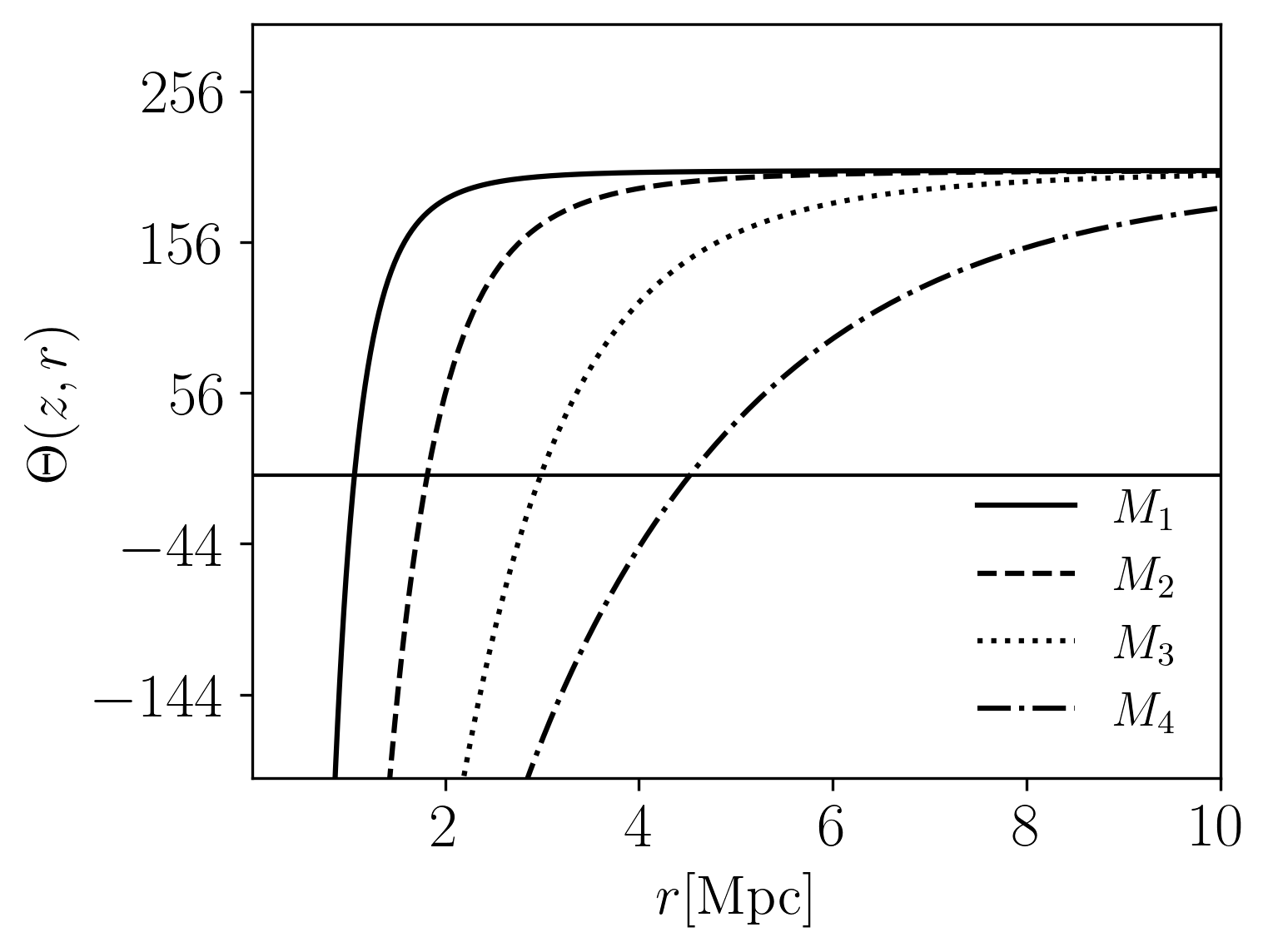}
\caption{\label{fig:local_group}  In the left panel,  we show the plot of ${\d \ln \rho}/{\d \ln r}$ vs the comoving radius. The position of the sharpest drop in density indicates the location of the halo boundary otherwise known as the splashback radius.   On the right panel, we show the expansion as a function of the comoving radius. The thick horizontal line corresponds to $\Theta = 0$. 
We considered the following halo masses 
$\left\{ M_1, M_2, M_3,M_4\right\} = \left\{ 1\times 10^{11},1\times 10^{12}, 1\times 10^{13},1\times 10^{14}\right\} M_{\otimes}$ and fixed the halo concentration to 
$c_{\rm{vir}} = 7$.   $M_{\otimes}$is the mass of the sun.
Note that the causal horizon is much greater than the splashback radius. Both of these radii can in principle be measured. 
  }
\end{figure}

This dichotomy may be better understood by calculating $\Theta$ in a perturbed FLRW spacetime  in comoving synchronous gauge 
\begin{eqnarray}\label{eq:synchronous_metric}
\d s^2  = -\d\tau^2 + h_{ij} \d x^i \d x^j\,,~~~~~{\rm{with}} ~~~~~~~~  h_{ij} = a^2\left[\left( 1 -2\psi\right)\delta_{ij}+\partial_{i}\partial_j E\right],
 \end{eqnarray}
 where  $\psi$ and $E$ are perturbed metric variables respectively.  Note that we neglect the vector and tensor perturbations since their propagation is null-like.  To the leading order, $\Theta$ is given by  
  \begin{eqnarray}
  \Theta(\tau,{\x}) \simeq 
  3 H(\tau) - \partial_{i}\partial_j \dot{E}(\tau,{\x})  =  \Theta_{H}+ \Theta_{L} \,,
  \end{eqnarray}
  where $``\,\dot{}\,"$ is the derivative with respect to proper time, $\Theta_H  = 3 H(\tau) $ and  $\Theta_{L}= - \partial_{i}\partial_j \dot{E}(\tau,{\x})$. Using the Poisson equation, it is possible to relate $\partial_{i}\partial_j E' $ to the matter density field $ \partial_{i}\partial_j E'  \propto -\delta'_{m}(\tau, {\r})$ leading to $\Theta_{L}\approx\delta'_{m}(\tau, {\r}) =- f(\tau) H(\tau) \delta_{m}( {\r})$~\cite{Villa:2015ppa}. Again, the expansion or contraction of nearby geodesics is determined by the relative dominance of  $\Theta_H$ and  $\Theta_{L}$.  
Within the Halo model~\cite{Cooray:2002dia}, it is possible to estimate ${\x}_{\rm{dp}}$ assuming  spherical symmetry, i.e ${\x}_{\rm{dp}} = r_{\rm{dp}} {\hat{r}}$ (${\hat{r}}$ is a unit vector).   So we can express the time derivative in terms of the radial derivative 
\begin{eqnarray}\label{eq:chain_rule}
 \frac{\partial \delta_{m}}{\partial \tau} = \frac{\partial \delta_{m}}{\partial r}  \frac{ \partial r}{\partial\tau} \,.
\end{eqnarray}
On the FLRW background spacetime, ${ \partial r}/{\partial\tau}$ can easily be evaluated for null geodesics. In the spherically symmetric  LTB model,
 there exists a clear relationship between the time and the areal radius~\cite{Clarkson:2010uz}. This is a general property of inhomogeneous spacetime. 
However, we consider a much-simplified approach  since the vanishing of $\Theta(\tau ,{\x})) $ is time independent, we can estimate  ${ \partial r}/{\partial\tau}$ using  the null geodesic relation. In this limit $r_{\rm{dp}}$ is interpreted as the comoving distance to the zero-velocity surface for an observer centred at $r=0$.  Implementing this in the equation 
  \eqref{eq:chain_rule} gives
\begin{eqnarray}\label{eq:expansion_halo}
\Theta(\tau, {\x})
\approx3 {H}(\tau) +  \frac{c}{r} \frac{\d \ln \rho}{\d \ln r} \,,
\end{eqnarray}
where $r$ is the comoving radial distance,  $c$ is the speed of light,
$\delta_{m}  \equiv {\delta \rho}/{\bar{\rho}} = ({\rho - \bar{\rho}})/{\bar{\rho}}\,$  and $\rho$ is the matter density. 
 $\Theta(\tau,{r}_{\rm{dp}})  =0$,   when
$
{r}_{\rm{dp} }= - { c} {\d \ln \rho}/({ 3 H}{\d \ln r}). 
$
Given any halo density profiles, it is straightforward to calculate ${\d \ln \rho}/{\d \ln r}$~\cite{Diemer:2017bwl}.  The plot of  $\Theta$ as function of $r$ is  shown in figure \ref{fig:local_group},   
We made use of the  NFW (Navarro–Frenk–White) dark matter density profile with the FLRW exterior~\cite{Diemer:2017uwt}.

There are three essential parts of $\Theta$ according to figure \ref{fig:local_group}: 

\begin{itemize}
\item Expanding region(+): There are regions with a comoving distance greater than the causal limit $ r>r_{\rm{dp}}$ for a given gravitationally bound cluster.  Within this region, the global Hubble expansion of spacetime dominates.  A typical example is void. 

\item  Collapsing region (-): These are regions ( $r<r_{\rm{dp}}$)  that have gravitationally decoupled from the Hubble expansion because they are moving with a slower velocity to catch up with the Hubble expansion. Within this region, a one-parameter family of nearby geodesics are collapsing with respect to an observer in the expanding region. Also with respect to the observer in the expanding region, the geodesics within this region would appear to be converging to caustics. 
%hence, the  $r <R$ region are not geodesic.

\item  Boundary:  This is a thin shell located at  $r =r_{\rm{dp}}$. Within the spherical collapse model, It is related to the turn-around radius. It is the critical point of equation \eqref{eq:volumeelement}. We refer to it as the causal horizon for massive particles with velocities less than three times the global Hubble rate.  Analysis of several observations and the N-body simulation of the local universe indicates that this scale exists and it is fundamental~\cite{Zuercher:2018prq,Murata:2020enz,Moon:2022nfg}. %Most discussion on this later. 

\end{itemize}

Finally, one key point to note here is that at $\Theta(\tau,{r}_{\rm{dp}})  = 0$, the determinant of the Jacobian is a non-zero constant. In the next sub-section, we study the dynamics of a one-parameter family of time-like geodesics in the neighbourhood of the causal horizon. 

\subsection{Caustics and inverse function theorem }\label{sec:piecewise_geodesic}

The existence of the causal horizon, i.e $\Theta(\tau,{r}_{\rm{dp}})  =0$ divides a family of time-like geodesics that starts at the same time in the past into two regions.  At $\tau = \tau_{\rm{ini}}$ hypersurface, one-parameter family of geodesics within $r <{r}_{\rm{dp}})$ are converging  while one-parameter family of geodesics in $r >{r}_{\rm{dp}}$ are expanding. 
In this sub-section, we study the dynamics of a one-parameter family of time-like geodesics in the neighbourhood of the causal horizon by perturbing the geodesics around $\Theta(\tau,{r}_{\rm{dp}})  =0$ surface:  
 \begin{eqnarray}
 \tau = \tau'  + \Delta \tau= \tau'  + \epsilon\,,
 \end{eqnarray}
 % \cite{Springel:2005mi,Adamek:2014xba}
  where $ \Delta \tau$ is an infinitesimally small difference between $\tau$ and $ \tau' $.  Under the infinitesimal perturbation, the Jacobian and the  expansion scalar change according to 
  \begin{eqnarray}
{\rm{det}[J]}(\tau,  {\x}_{\rm{dp}})  &=& {\rm{det}[J]}(\tau, {\x}_{\rm{dp}}) + \frac{\d{\rm{det}[J](\tau,{\x}_{\rm{dp}})}}{\d \tau}\bigg|_{\tau = \tau'} \Delta \tau+ \mathcal{O}(\Delta \tau)^2\,,
\\
  \Theta(\tau, {\x}_{\rm{dp}})& =& \Theta(\tau', {\x}_{\rm{dp}}) + \frac{\d \Theta(\tau,{\x}_{\rm{dp}})}{\d\tau} \bigg|_{\tau = \tau'}\Delta \tau+ \mathcal{O}(\Delta \tau)^2\,.
 \end{eqnarray}
Substituting  in equation \eqref{eq:volumeelement}  and keeping only terms that are of linear order in $\Delta\tau$ gives
\begin{eqnarray}
  \frac{\d^2 {\rm{det}[J]}(\tau,{\x}_{\rm{dp}})}{\d \tau^2}\bigg|_{\tau = \tau'} &=&  {\rm{det}[J]}(\tau,{\x}_{\rm{dp}})  \frac{\d \Theta}{\d \tau}\bigg|_{\tau = \tau'} + \mathcal{O}(\Delta \tau)^2\,.
\end{eqnarray}
 The time evolution of $\Theta$ is given by the Raychaudhuri equation \eqref{eq:expansion} and imposing  $ \Theta(\tau',{\x}_{\rm{dp}})= 0$,  leads to 
 \begin{eqnarray}
 \frac{1}{ {\rm{det}[J]}(\tau,{\x}_{\rm{dp}}) }   \frac{\d^2 {\rm{det} [J](\tau,{\x}_{\rm{dp}})}}{\d \tau^2}\bigg|_{{\tau}={\tau'}}&=&
 - \left[ {\sigma}_{ab}{\sigma}^{ab}  + {R}_{ab} {u}^a {u}^b\right] <0\,.
 \label{eq:translation2}
\end{eqnarray}
Assuming that the weak energy condition holds $ R_{\alpha\beta}  u^{\alpha} u^{\beta}  \ge 0$ and we know that ${\sigma}_{ab}{\sigma}^{ab}$ is positive definite, then the second derivative must be negative indicating  that ${x}_{\rm{dp}}$  is a local maximum. This is a typical example of a ball rolling down a hill.  Any slight perturbation in the particle position causes it to roll down hill.
The ball accelerates as it rolls down the hill because the gravitational force is pulling it downwards.

Most crucial  lesson here is that the global Hubble expansion breaks down at ${\x}_{\rm{dp}}$, therefore, an expanding coordinate system cannot be extended  beyond the zero-velocity surface.   Extending the geodesics that started out in an expanding spacetime beyond ${\x}_{\rm{dp}}$ will end up in a singularity or a caustic immediately after ${\x}_{\rm{dp}}$, hence ${\x}_{\rm{dp}}$ is a boundary. 
The fact that the determinant of the Jacobian ($ {\rm{det}}[J](\tau, {\x})$) is a non-zero constant at ${\x}_{\rm{dp}}$ provides hints on how to proceed. For example,  by the inverse function theorem, it indicates that we can find another more suitable set of  coordinates in the immediate neighbourhood of ${\x}_{\rm{dp}}$ on the collapsing region and join it seamlessly to the FLRW spacetime at ${\x}_{\rm{dp}}$.  We show in section \ref{sec:vorticit_generation} that conditions for joining two families of geodesics across the zero-velocity surface. 

% leads to a generation of vorticity which then alters the key assumption of the focussing theorem, thereby avoiding the singularity. 

\section{Model of the universe with collapsing and expanding regions}\label{sec:universe_model}
%\ref{subsec:separate_universe} and

 The analysis in sub-section \ref{sec:piecewise_geodesic} shows that the one-parameter family of geodesics which describes the dynamics of massive particles in the expanding regions of the universe cannot be extended beyond the causal horizon. 
 At the causal horizon, the Jacobian determinant is constant and by inverse function theorem, we can define another one-parameter family of geodesics to describe the dynamics in the collapsing region since the extension of  the geodesics in the expanding universe into the collapsing region leads to a caustics.  We describe in detail how to do this in the remainder of this section.

\subsection{Dynamics of geodesics in both regions and their junction conditions }\label{sec:junction_conditions}

  The diffeomorphism symmetry of general relativity in 4 dimensions allows freedom to choose coordinates.
We want to find a smooth coordinate transformation that will smoothly join the expanding and collapsing coordinates. Therefore, we  require that the spatial coordinates satisfy the following condition at the boundary
\begin{eqnarray}\label{eq:coordinate_trans}
 {x}^{a}_{-} |_{\Sigma}= x_{+}^{a}|_{\Sigma} \,,
\end{eqnarray}
where $x_{+}^{a} $ is the coordinates in the expanding region, ${x}^{a}_{-} $ is the coordinates in the collapsing region and ${\Sigma}$ is the spatial hyoersurface.
We parameterise the geodesics by a translated time parameter $t   = \tau -\tau_{\rm{ini}}(r_{\rm{dp}})$, where $\tau_{\rm{ini}}(r_{\rm{dp}})$,  perhaps is related to the bang time when collapsing and expanding regions of the universe were delineated. This parameterisation allows us to place the boundary hypersurface $\Sigma$ at $t =0$. Thus, geodesics with positive $t >0$ ($ \tau >\tau_{\rm{ini}}(r_{\rm{dp}})$) are in the expanding spacetime while the geodesics with $t <0$ ($ \tau <\tau_{\rm{ini}}(r_{\rm{dp}})$) are in the collapsing spacetime. 
Therefore, we define the 4-velocities in the collapsing and expanding spacetimes as 
  \begin{eqnarray}
   u^a_{-}   = \frac{\d x^a_{-}}{\d t}  ~~~~{\rm{and}}~~~~~~~u^a_{+}
 =\frac{\d x^a_{+}}{\d t} 
  \end{eqnarray}
  respectively. 
  Note that these vectors are time-like $u^a_{\pm} u_{a}^{\pm} = -1$ in there respective regions. 
  The two regions are modelled as oriented Lorentzian manifolds denoted as $\mathcal{M}_{\pm}= \mathcal{M}_{\pm} \cup \partial \mathcal{M}_{\pm}$.  The boundary lies on the hypersurface $\Sigma$ of both spacetimes $\Sigma \in \partial \mathcal{M}_{\pm}$. We consider a situation where both spacetimes can be combined into an ambient spacetime ($M,g$), whose manifold is the union of the manifolds of the individual parts such that $\mathcal{M}_{4} = \mathcal{M}_{-} \cup  \mathcal{M}_{+}$.
  
  This setup is better understood using the language of the distribution function.  The Heaside function $\mathbb{H}( t)$ is used to constrain evolutionary history in both manifolds. The Heaside function is normalised such that it is equal to $+1$ if $ t >0$, $0$ if $ t<0$ and indeterminate if $ t = 0$, with the following properties 
\begin{eqnarray}
\mathbb{H}^2( t) = \mathbb{H}( t)\,, \qquad \mathbb{H}( t) \mathbb{H}(- t) = 0\,, \qquad \frac{\d }{\d  \ell} \mathbb{H}( t) = \delta ( t)  \,,
\end{eqnarray}
where $\delta( t)$ is the Dirac distribution. The metric of the ambient spacetime is related to metrics in  $\mathcal{M}_{+}$ and $\mathcal{M}_{-}$ as 
\begin{eqnarray}
g_{ab} = \mathbb{H}(- t)g^{-}_{ ab}   +  \mathbb{H}( t)\ g^{+}_{ab} + \delta(t) \delta g_{ab}\,,
\end{eqnarray}
where  last term $\delta g_{ab}$ denotes the metric at the boundary. For the smooth joining of the metrics at the boundary, we require that $\delta g_{ab} = 0$ vanishes and the metrics join smoothly at the boundary~\cite{Israel:1966rt}
\begin{eqnarray}
g^{-}_{ ab}\big|_{\Sigma}  = g^{+}_{ab} \big|_{\Sigma}\,.
\end{eqnarray}
In order to reduce clutter, we drop $ \mathbb{H}$ in the subsequent discussion. 
We define the action of the curve between point $p$ to point $q$ in the ambient spaceetime as a sum of the actions of the smooth curves in the two manifolds 
\begin{eqnarray}\label{eq:oiecewise_action}
S &=& S_{-} + S_{+} =  \int_{p}^{ t_{\rm{dp}}} {L} _{-}\left[ \gamma_{-}( t) , \gamma'_{-}( t)\right] \d  t+ 
 \int_{ t_{\rm{dp}}}^{q} {L} _{+}\left[ \gamma_{+}( t) , \gamma'_{+}( t)\right] \d  t\,,
\end{eqnarray}
where $L_{-}$ is the Lagrangian for the smooth curves in the collapsing region and $L_{+}$ is the Lagrangian for smooth curves in the expanding region. %We require that the curves join smoothly at $ t_{\rm{dp}}$. 
Prime indicates a derivative with respect to the argument.  For a smooth curve  $\gamma_{-}$ is geodesic within the range $[p,t_{\rm{dp}}]$ and  $\gamma_{+}$ is geodesic within the range $[t_{\rm{dp}},q]$. 
%Equation \eqref{eq:oiecewise_action} indicates that we can describe the total action for the geodesic as a sum of two geodesics. 
The critical point of the total action with respect to the infinitesimal variation as described in sub-section \ref{sec:Geodesic_equation}  corresponds to 
\begin{eqnarray}\label{eq:vary_action}
0  = \frac{\d S }{\d s} \bigg|_{s = 0}&=& \frac{\d S _{-}}{\d s} \bigg|_{s = 0}+ \frac{\d S _{+}}{\d s} \bigg|_{s = 0}
\\
&=&\frac{\d }{\d s} \bigg|_{s = 0} \int_{p}^{t_{\rm{dp}}}L _{-}\left[\gamma_{-}( t),{{\gamma}'_{-}}( t)\right] \d  t  + 
\frac{\d }{\d s} \bigg|_{s= 0} \int_{ t_{\rm{dp}}}^{q}L _{+}\left[\gamma_{+}( t),{{\gamma}'_{=}}( t)\right] \d  t\,.
\end{eqnarray}
Then varying both actions  following the steps described in section \ref{sec:Geodesic_equation} without imposing the proper variation condition to zero gives 
\begin{eqnarray}
\frac{\d S_{-} }{\d s} \bigg|_{s = 0}&=& \frac{\partial L_{-}}{\partial {{\gamma'}^i_{-}} }( t^{-})\xi^i_{{-}}( t^{-}) \bigg|_{p}^{t_{\rm{dp}}}+ \int^{ t_{\rm{dp}}}_{p}  \left(\frac{\partial L_{-}}{\partial \gamma_{-}^i} - \frac{\d}{\d  t } \frac{\partial L_{-}}{\partial {{\gamma'}_{-}^i} }\right)\xi_{-}^i( t)  \d  t\ ,
\\
\frac{\d S_{+} }{\d s} \bigg|_{s = 0} &=& -  \frac{\partial L_{+}}{\partial {{\gamma'}_{+}^i} }( t^{+})\xi_{+}^i( t^{+})\bigg|_{t_{\rm{dp}}}^{q}+ \int^{q}_{ t_{\rm{dp}}}  \left(\frac{\partial L_{+}}{\partial \gamma_{+}^i}- \frac{\d}{\d  t } \frac{\partial L_{+}}{\partial {{\gamma'}_{+}^i} }\right)\xi_{+}^i( t)  \d  t\,.
\end{eqnarray}
Now imposing proper variation at the  endpoints of the geodesic $\xi_{-}^i(p) = \xi_{+}^i(q)$  and not at the boundary gives 
\begin{eqnarray}
0 &=& \left[ \frac{\partial L _{-}}{\partial {{\gamma'}^i_{-}} }( t_{\rm{dp}})\xi_{-}^i( t_{\rm{dp}}) - 
 \frac{\partial L _{+}}{\partial {{\gamma'}_{+}^i} }( t_{\rm{dp}})\xi_{+}^i( t_{\rm{dp}}) \right]
\\  \nonumber&&
~~~~~~~~~~~~~+ \int^{ t}_{p}  \left(\frac{\partial L_{-}}{\partial \gamma_{-}^i}( t) - \frac{\d}{\d  t} \frac{\partial L_{-}}{\partial {{\gamma'}_{-}^i} }\right)\xi_{-}^i(t)  \d  t
+\int^{q}_{t}  \left(\frac{\partial L_{+}}{\partial \gamma_{+}^i}( t) - \frac{\d}{\d  t} \frac{\partial L_{+}}{\partial {{\gamma'}_{+}^i} }\right)\xi_{+}^i( t)  \d  t \,.
\end{eqnarray}
Given the smoothness condition for the coordinates given in equation \eqref{eq:coordinate_trans}, we  impose that the curves are piecewise smooth, that is $\xi^i_{-}(t) = \xi_{+}^i (t)= \xi^i (t)$ at the boundary, therefore, the Euler Lagrange  equation holds in the separate spacetimes 
\begin{eqnarray}\label{eq:EL_A}
 \frac{\d}{\d  t} \frac{\partial L_{-}}{\partial {{\gamma'}_{-}^i} } -\frac{\partial L_{-}}{\partial \gamma_{-}^i}&=&  0 \qquad {\rm{for}}\quad   t \in [p, t_{\rm{dp}}) \,,
 \\
  \frac{\d}{\d  t} \frac{\partial L_{+}}{\partial {{\gamma'}_{+}^i} }-\frac{\partial L_{+}}{\partial \gamma_{+}^i} &=&  0 
  \qquad {\rm{for}} \quad  t \in ( t_{\rm{dp}},q] \,.
  \label{eq:EL_C}
\end{eqnarray}
Inserting the Lagrangian introduced in equation  \eqref{eq:length_functional} into equations  \eqref{eq:EL_A} and \eqref{eq:EL_A} gives the corresponding geodesic equations in both spacetime regions. %\eqref{eq:geodesic_equation} 
And at the boundary, we have the following condition that must hold
\begin{eqnarray}\label{eq:Israel}
 \left[ \frac{\partial L_{-}}{\partial {{\gamma'}_{-}^i} }( t_{\rm{dp}}) -  \frac{\partial L_{+}}{\partial {{\gamma'}_{+}^i} }(t_{\rm{dp}}) \right]\xi^i( t_{\rm{dp}})  = 0\,.
\end{eqnarray}
 It will become clearer shortly that equation \eqref{eq:Israel} is the generalised Israel junction conditions~\cite{Israel:1966rt}.  
 The boundary conditions for the 4-vector for the  geodesic equation are obtained by plugging in tthe geodesic equation given in equation \eqref{eq:length_functional}
\begin{eqnarray}\label{eq:four_velocity}
 \left[\frac{1}{L_{-}}\frac{\d x_{+}^a }{\d t}( t_{\rm{dp}}) - \frac{1}{L_{+}}  \frac{\d x_{+}^a }{\d t}( t_{\rm{dp}})  \right]\approx \left[\frac{1}{L_{-}}u^a_{-}\bigg|_{\Sigma}- \frac{1}{L_{+}} u^a_{+}\bigg|_{\Sigma}\right]\xi^i( t_{\rm{dp}}) = 0 \,.
\end{eqnarray}
The geodesic deviation equation associated with the piecewise geodesics can be obtained by performing a second variation of equation \eqref{eq:oiecewise_action} as discussed in sub-section \eqref{eq:GDE_variation}. However, since equations \eqref{eq:EL_A} and \eqref{eq:EL_C} are simply Euler-Lagrange equations of motion we can obtain the respective geodesic deviation equation using the Lagrangian for the geodesic deviation equation given in equation \eqref{eq:Lagrangian_GDE}.
Furthermore, putting the Lagrangian for the geodesic deviation equation given in equation \eqref{eq:Lagrangian_GDE} in the generalised Junction condition gives the junction condition for the second fundamental form
%the second Israel Junction condition~\cite{Israel:1966rt}
\begin{eqnarray}\label{eq:Second_fundamental_form}
\left[\frac{\d \xi^a}{\d  t}( t_{\rm{dp}})  - \frac{\d \xi^a}{\d  t}( t_{\rm{dp}})\right] \xi_{a} = 0 = \left[B_{ab}\bigg|_{\Sigma} - B_{ab}\bigg|_{\Sigma}\right]\xi^{a} \xi^{b} = \left[K_{ab}^{-} - K_{ab}^{+}\right]\xi^{a} \xi^{b}\,,
\end{eqnarray}
where we made use of equation \eqref{eq:firstorderode} in the third equality.  
Equations \eqref{eq:four_velocity} and \eqref{eq:Second_fundamental_form}  define the junction conditions that allow glueing spacetimes together across a boundary hypersurface $\Sigma  =  \partial M_{+} \cap \partial M_{-}$ via a thin shell.  Note that equation \eqref{eq:Second_fundamental_form} is the second Israel Junction condition~\cite{Israel:1966rt}.

\subsection{Gravity and the validity of fluid approximation}\label{sec:fluid_element}

We showed in the previous sub-section that the general conditions for joining two families of geodesics at a given boundary. We derived the junction conditions that the 4-velocity vector and the second fundamental form that the two families of geodesics must satisfy.
%All of these derivations simply imposed general energy conditions on the matter content of the universe.  
We have not explicitly made use of any specific theory of gravity. What we have derived so far will apply to any theory that respects the principle of least action.  Now, we need to derive corresponding conditions for a given theory of gravity.  Here, we consider the Einstein general theory of  relativity, it relates the geometry of spacetime  to the matter content of the universe:  
\begin{eqnarray}\label{eq:GR}
 G_{ab} = {R}_{ab} - \,\frac{1}{2}{R}\,{g}_{a b} =   {T}_{a b} \,,
\end{eqnarray}
 where $T^{ab}$  is the energy-momentum tensor and $R$ is the Ricci scalar.  We have so far derived the generalised junction conditions and equation of motion for the trajectory of massive particles on curved spacetime.  Equation \eqref{eq:GR} may be obtained from the Einstein-Hilbert action. For oriented manifolds such as those described in sub-section \ref{sec:junction_conditions},  the Einstein-Hilbert action has non-vanishing boundary terms~\cite{York:1972sj,Gibbons:1976ue}. 
  One could vary the Einstein-Hilbert action similarly as we did in equation \eqref{eq:vary_action} and avoid setting the tangential derivatives to zero at the boundary~\cite{poisson_2004} to obtain the corresponding equation of motion. 
  A slightly different way to obtain the same result is to recall that Einstein's theory of gravity is a second-order partial differential equation. In this approach, one finds that the momentum constraint component of the Einstein field equations contains the tangential derivative of the extrinsic curvature tensor which does not need to vanish at the boundary.  This constraint can easily be derived using the Gauss-Codazzi identity and equation \eqref{eq:four_velocity}. With respect to the Einstein field equation, this is interpreted as stress-energy tensor at the boundary~\cite{Goldwirth:1994ru}
\begin{eqnarray}\label{eq:boundary_stress}
\mathcal{S}_{ab} = - \frac{1} {8\pi } \left( \left[[K_{ab} \right]] - \left[[K \right]]h_{ab} \right)  \,, ~~~~~{\rm{where}}~~~~~\left[[K_{ab} \right]]  = K_{ab}^{-} - K_{ab}^{+} ~~~~{\rm{and}} ~~~ K = h^{ab}K_{ab} \,.
\end{eqnarray}
Note that the violation of the $\doublesqbraket{ K_{ab}} = 0$ implies that the spacetime is not smooth at $\Sigma$. This has sound physical interpretation because it indicates that the surface layer has a non-vanishing stress-energy tensor.  In cosmology for Gaussian initial conditions, one expects fluctuations of order $10^{-5}$ in the matter density field on large scales. On small scales, the fluctuations are much larger.  
   \begin{figure}[h]\label{fig:fluid_approximation}
  \includegraphics[width=140mm,height=80mm] {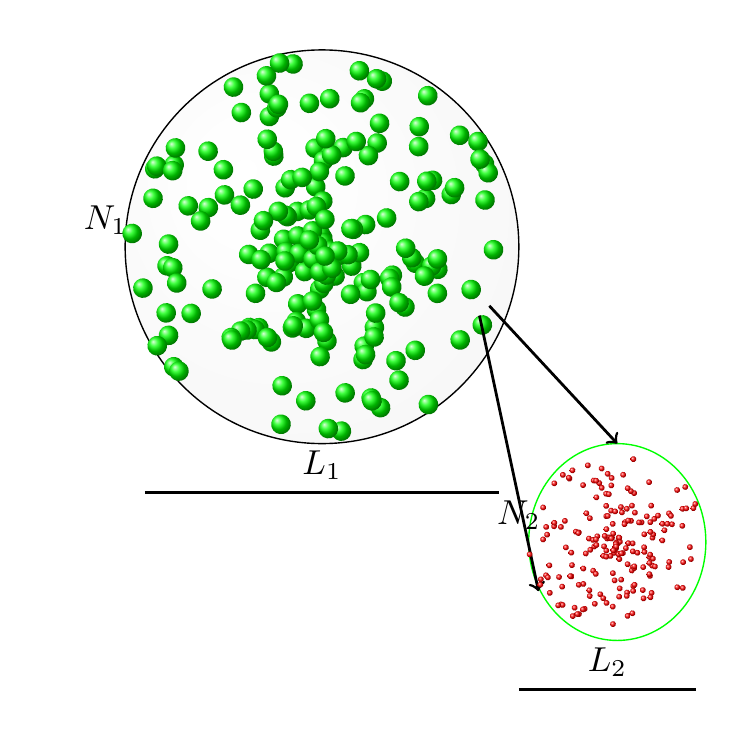}
 \caption{The observed universe with a characteristic size $L_{1}$ is modelled as a fluid that contains $N_1$ fluid elements, i.e the green balls. The green balls are gravitationally bound clusters, they are virialised.  The fluid description description of the  evolution of the universe breaks down when the interaction length between the fluid elements is of the order of $L_{2}$.  That is the condition for validity of fluid approximation is given by $N_1 \gg N_2 \gg1$ and $L_1\gg L_2$. We can extend this analogy because in cosmology  the red balls are not fundamental particles. We consider them as stars, hence we can also describe the dynamics within the green ball using fluid approximation where the red balls are fluid elements.
  }
 \end{figure}
 
 In physics, the elementary particles of nature are leptons, quarks, and gauge bosons. These particles are quantum mechanical in nature. It is not clear yet how to fit quantum mechanism and general relativity together. So we cannot associate elementary particles to the trajectory of massive particles we have derived so far. 
 In cosmology, however, fundamental interactions between these elementary particles are not important,  rather the dynamics of planets,  stars, galaxies, clusters, etc are important depending on the length scale of interest.
On the Giga-parces scales, for example, one could consider, galaxies as fluid parcels or fluid elements and assign each fluid parcel a geodesic. This is known as a fluid approximation. It is assumed that on Giga-parces scales, the internal dynamics of a galaxy are not important, hence, interactions within it are averaged over.  
Even in the N-body simulation in cosmology, a similar approximation is made but it is interpreted as mass resolution. In this case, a fluid element which in principle is made up of many dark matter particles is assigned a mass and a geodesic~\cite{Navarro:1995iw}. The N-body simulation evolves the fluid element and not individual dark matter particles.

The fluid approximation  breaks down when the internal dynamics within the fluid element become important.
In our case with clusters as fluid elements, the fluid approximation will break down at the causal limit or the zero-velocity surface. 
%This is also the scale where the one-parameter family of geodesics breaks down(see sub-section \ref{subsec:separate_universe}).  
This indicates that internal dynamics of clusters cannot be ignored. Therefore, in order to describe what happens within clusters, we could consider stars as fluid elements and assign a different one-parameter family of geodesics to each state.  The fluid approximation will apply but in a different one-parameter family of geodesics.  The conditions for the validity of fluid approximation are described in detail in figure \ref{fig:fluid_approximation}. The essential point is that fluid element before shell crossing is different from  the fluid element after shell crossing. 
\subsection{ Fluid rest frame and the far-away observer 4-velocity}

In the end, the total energy-momentum tensor in the ambient spacetime would include the stress-energy tensor at the boundary of the different fluid approximations
\begin{eqnarray}
T^{\rm{tot}}_{ab} = T^{+}_{ab} +  T^{-}_{ab} +  \mathcal{S}_{ab}\,,
\end{eqnarray}
where $T^{+}_{ab}$ and $T^{-}_{ab}$  are the energy-momentum tensor in the expanding and collapsing regions. $\mathcal{S}_{ab}$ is the stress-energy momentum tensor due to a jump discontinuity in the Riemann tensor. 
The physical interpretation of $\mathcal{S}_{ab}$ is given in terms of the energy-momentum tensor.  Within the standard cosmological model, for example, the late universe is dominated by the cosmological constant and dust. In this limit $T^{+}_{ab}$ may be decomposed as 
\begin{eqnarray}\label{eq:emt_decomp}
T^{ab}_{+} = \tilde{\rho}_{m+} \tilde{u}^a_{+} \tilde{u}^b_{+}\,,  \qquad \qquad T^{ab}_{-} = \tilde{\rho}_{m-}  \tilde{u}^a_{-}  \tilde{u}^b_{-} \qquad {\rm{and}}\qquad \doublesqbraket{T^{ab}}  = 0\,,
\end{eqnarray}
where $\tilde{\rho}_{m\pm} $ is the matter density field $\tilde{\rho}_{m \pm}  ={T}_{{\pm} ab}\tilde{u}^a_{\pm}\tilde{u}^b_{\pm}\,$.  Note that one can work with perfect fluid or fluid with non-vanishing anisotropic stress.
For the stress-energy tensor due to the jump in the Riemann tensor, it can also be decomposed in a similar way \cite{poisson_2004}
\begin{eqnarray}\label{eq:shell_crossing_EMT}
\mathcal{S}_{ab} %= - \frac{1} {8\pi} \left( \left[[K_{ab} \right]] - \left[[K \right]]h_{ab} \right) 
=  - \frac{1} {8\pi}
 \bigg[ \left({K}_{+ ab}-{K}_{-ab}\right) - \left({K}_{+}- {K}_{-}\right) h_{ab}\bigg] = \mathcal{S}_{+ab} -\mathcal{S}_{-ab} \,.
\end{eqnarray}
Note that the extrinsic  curvature is related to the covariant derivative of $K_{ab}= h_{a}{}^{c}\nabla_{b}\tilde{u}_{Sc}$
\begin{eqnarray}\label{eq:matter_decomp}
 \nabla_{b}\tilde{u}_{\mathcal{S}a}=-\tilde{u}_{\mathcal{S}b} \tilde{A}_{\mathcal{S}a}+\frac{1}{3}\tilde{\Theta}_{\mathcal{S}} \tilde{h}_{\mathcal{S}ab }
+\tilde{\sigma}_{\mathcal{S}ab} +\tilde{ \omega}_{\mathcal{S}ab} \,,
\end{eqnarray}
where $\tilde{A}_{\mathcal{S}a} = \tilde{u}^b_{\mathcal{S}} \nabla_b \tilde{u}_{\mathcal{S}a}$ is the acceleration in the rest frame of ${u}_{Sa}$, $\tilde{\Theta}_{\mathcal{S}}$ is expansion of the  geodesics $\tilde{u}_{Sa}$, $\tilde{\sigma}_{\mathcal{S} ab} $ and $\tilde{ \omega}_{\mathcal{S} ab}$ are the corresponding shear and vorticity respectively.

Furthermore,  it is instructive to interpret contributions to $\mathcal{S}_{ab}$ (i.e equation \eqref{eq:shell_crossing_EMT}) in a similar fashion as $T^{ab}_{\pm}$.  This can be done by decomposing $\mathcal{S}^{ab}$ into irreducible units with respect to $\tilde{u}^a_{\mathcal{S}}$  and $\tilde{h}_{\pm ab}$%in a similar form as as equation \eqref{eq:emt_decomp} using equation \eqref{eq:matter_decomp}
\begin{eqnarray}\label{eq:Sab}
\mathcal{S}^{ab} = \tilde{\rho}_{\mathcal{S}} \tilde{u}^a_{\mathcal{S}} \tilde{u}^b_{\mathcal{S}} + \tilde{P}_{\mathcal{S}}\tilde{h}^{ab}_{\mathcal{S}}
  + \tilde{q}^{(a}_{\mathcal{S}} \tilde{u}^{b)}_{\mathcal{S}} + \tilde{\pi}^{\<ab\>}_{\mathcal{S}} \,,
\end{eqnarray}
where $ \tilde{\rho}_{\mathcal{S}} $, $\tilde{P}_{\mathcal{S}}$, $\tilde{q}_{\mathcal{S} a}$ and $\tilde{\pi}^{\<ab\>}_{\mathcal{S}}$ are the corresponding boundary layer energy density, pressure, energy flux vector and  stress-energy tensor respectively:
\begin{eqnarray}
\tilde{\rho}_{\mathcal{S}}  =\mathcal{S}_{ab}\tilde{u}^a_{\mathcal{S}} \tilde{u}^b_{\mathcal{S}}\,, \qquad \tilde{P}_{\mathcal{S}}  =\frac{1}{3} \tilde{h}_{\mathcal{S}ab} \mathcal{S}^{ab}\,, \qquad \tilde{q}_{\mathcal{S} \<b\>} = - \mathcal{S}_{a\<a\>} \tilde{u}^a_{\mathcal{S}}\,, \qquad  \tilde{\pi}_{\mathcal{S}\<ab\>} = \mathcal{S}_{\<ab\>} \,.
\end{eqnarray}
Using equation \eqref{eq:matter_decomp} in equation \eqref{eq:shell_crossing_EMT}, we can calculate these observables
\iffalse
\begin{eqnarray}
\tilde{\rho}_{\mathcal{S} }  &=& 0\,, ~~~~~~~~~~~~~~~~~~~
\tilde{P}^{\mathcal{S}} =  \frac{1} {2\pi}\frac{1}{3}\left[\tilde{\Theta}^{\mathcal{S}}\right] \,, %=P_{+\mathcal{S}} -P_{-\mathcal{S}}\,,
\label{eq:pressure_def}%= \frac{\varepsilon} {8\pi} 
\\
\tilde{q}^{\mathcal{S} }_{\<a\>}  &=& -\frac{1} {8\pi} \left[\tilde{A}^{\mathcal{S}}_{\<a\>}\right]\,,
~~~~~~~~
 \tilde{\pi}^{\mathcal{S}}_{\<ab\>} =-\frac{1} {8\pi}\left[ \tilde{\sigma}^{\mathcal{S}}_{ab} - \frac{2}{3}\tilde{h}^{\mathcal{S}}_{ab} \tilde{ \Theta}^{\mathcal{S}} \right] \,.
\end{eqnarray}
\fi
\begin{eqnarray}
\tilde{\rho}_{{\mathcal{S}} }  &=& 0\,, ~~~~~~~~~~~~~~~~~~~
\tilde{P}_{\mathcal{S}}  =  \frac{1} {2\pi}\frac{1}{3}\doublesqbraket{\tilde{\Theta}_{\mathcal{S}} }\,, %=P_{+\mathcal{S}} -P_{-\mathcal{S}}\,,
\label{eq:pressure_def}%= \frac{\varepsilon} {8\pi} 
\\
\tilde{q}_{S\<a\>}  &=& -\frac{1} {8\pi} \doublesqbraket{\tilde{A}_{ \<a\>}}\,,
~~~~~~~~
 \tilde{\pi}_{S\<ab\>} =-\frac{1} {8\pi}
 \doublesqbraket{ \tilde{\sigma}_{ab} - \frac{2}{3}\tilde{h}_{ab}  \tilde{\Theta} }\,.
\end{eqnarray}
As we shall see later, these fluid variables (i.e. $\tilde{P}_{\mathcal{S}}$, $\tilde{q}_{\mathcal{S} a}$ and $\tilde{\pi}^{\<ab\>}_{\mathcal{S}}$) are generated by the relative motion between  adjacent fluid elements in the neighbourhood of the boundary. The relative motion induces
internal friction (viscosity) at the boundary even if it were a perfect fluid in the bulk.  
To capture this effect, we parametrise
 $\pi^{\<ab\>}_{+ } $ in terms of  the bulk and shear viscosity components
\begin{eqnarray}
\tilde{\pi}^{\<ab\>}_{+ } = -\left[  \zeta_{+} \tilde{\Theta}_{+}  h^{ab}_{+}  +2\eta_{+} \tilde{\sigma}^{ab}_{+} \right]\approx -2\eta_{+} \tilde{\sigma}^{ab}_{+} \,,~~~~~~~\tilde{\pi}^{\<ab\>}_{- } = -\left[  \xi_{-} \tilde{\Theta}_{-}  \tilde{h}^{ab}_{-}  +2\eta_{-} \tilde{\sigma}^{ab}_{-} \right]\approx-2\eta_{-} \tilde{\sigma}^{ab}_{-}\,,
\end{eqnarray}
 where $\xi_{\pm} > 0$ and $\eta_{\pm} >0$ are bulk and shear viscosity respectively. In the second equality, we have approximated $\tilde{\pi}^{\<ab\>}_{\pm } $ with the shear viscosity only for simplicity.  
 It is straightforward, to include the additional contribution due to the bulk viscosity. %Note that $\mathcal{S} = \{+,-\}$. 

%\begin{eqnarray}\label{eq:EMT2}
%{T_{\rm{tot}}}_{+ab}& =&  {\rho}_{m+}  {u}^{+}_a{u}^{+}_b +P_{+}{h}_{+}^{ab}  + {q}_{+}^{(a} {u}^{b)}_{+} - 2\eta_{+} {\sigma}^{ab}_{+}\,,
%&\approx& {\rho} {u}_a{u}_b   + {q}_{S}^{(a} {u}^{b)}  - 2\eta {\sigma}^{ab}_{S}\,.
%\end{eqnarray}
%A similar expression exists for ${T_{\rm{tot}}^{-}}_{ab}$.
%These fluid 4-velocity are defined in the rest frame of the fluid, however, what is really important to us is how a far-away observer sees the fluid flow between the fluid elements in both regions of the spacetime. 
%\subsection{ Fluid rest frame and the far-away observer 4-velocity}

The energy-momentum tensor for each of the fluid specie is measured in its rest frame is given by  
% \begin{eqnarray}\label{eq:effective_emt}
%{T}_{+\rm{tot}}^{ab} &=& T^{+}_{ab} +  T^{-}_{ab} +  \mathcal{S}_{ab}  = 
%\tilde{\rho}_{m+} \tilde{u}^a_{+}\tilde{u}^b_{+} +  \tilde{P}_{+}\tilde{h}^{ab}_{+} + \tilde{q}^{(a}_{+} \tilde{u}^{b)}_{+}   - \xi_{+} \tilde{\Theta}_{+} \tilde{h}^{ab}_{+}- 2 \eta_{+} \tilde{\sigma}^{ab}_{+}\,,
%\end{eqnarray}
 \begin{eqnarray}\label{eq:effective_emt}
\tilde{T}_{\rm{tot}}^{ab} = \tilde{\rho}_{m+} \tilde{u}^a_{+}\tilde{u}^b_{+} + \tilde{\rho}_{m-} \tilde{u}^a_{-}\tilde{u}^b_{-}  +  \left[\tilde{P}_{\mathcal{S} }\tilde{h}^{ab}  
+ \tilde{q}^{(a}_{{\mathcal{S}}} \tilde{u}^{b)}_{\mathcal{S}} - 2\eta \tilde{\sigma}^{ab} \right]\,,
\end{eqnarray}
where we have replaced $\mathcal{S}_{ab}  $ with
\begin{eqnarray}
\mathcal{S}_{ab}  &=& \tilde{P}_{\mathcal{S} }\tilde{h}^{ab}  
+ \tilde{q}^{(a}_{{\mathcal{S}}} \tilde{u}^{b)}_{\mathcal{S}} - 2\eta \tilde{\sigma}^{ab}  \,.%\delta^{\Sigma}\,,
\end{eqnarray}
%where we made use of a tilde to indicate quantities measured in the rest frame of the fluid.  
The  energy-momentum tensor that goes into the Einstein field equation in the ambient spacetime is the sum of the energy-momentum tensor  of the fluid element decomposed in terms of the threading 4-vector. The matter 4-velocity is related to the fundamental(threading time-like) 4-velocity (we  refer to this as the 4-velocity of the observer)  according to 
\begin{eqnarray}\label{eq:relativevelocity}
\tilde{u}^a_{+}=\gamma\left({u}^a_{+}+v^a_{+}\right) \approx  {u}^a_{+}+v^a_{+}\,,\quad v^a_{+}{u}^{+}_a=0  \quad {\rm{and}}\quad \gamma_{+}={{(1-v^2_{+})}}^{-\frac{1}{2}} \,,
\end{eqnarray}
where $v^a_{+}$ is the relative velocity between the matter and observer frames and  $\gamma_{+}$ is the Lorentz boost factor.   The projection tensor to matter hypersurface is given by $g^{+}_{ab} = h^{+}_{ab}- u^{+}_{a}u^{+}_{b} = \tilde{h}^{+}_{ ab}- \tilde{u}^{+}_{ a}\tilde{u}^{+}_{ b} $ and at the leading order in $v^a_{+}$ is given by
\begin{eqnarray}\label{eq:hypersurface_fluid}
\tilde{h}^{+}_{ab} &\approx& h^{+}_{ ab}+\left[2u^{+}_{(a}{v}^{+}_{b)}
+{v}^{+}_{a} {v}^{+}_{b}\right]  + \mathcal{O} \left(\epsilon v^2_{+}\right) \,.
\end{eqnarray}
The full covariant decomposition of spacetime covariant derivative of  $\tilde{u}^a_{+}$ is given in equation \eqref{eq:matter_decomp}.
% At the leading order in $v^a$, $\tilde{u}^a$ is related to $v^a$ according to $\tilde{u}^a = {u}^a+v^a$.
  The decomposition of the  full spacetime covariant derivative of $v^a_{+}$ with respect to $u^a_{+}$ is given by 
\begin{eqnarray}
\nabla_{a}v_{+b}&=&-\dot u_{+c}  v^c_{+}\, u_{+a} u_{+b} - u_{+a}\dot v_{+\langle b\rangle}
+\left( \frac{1}{3}\Theta_{+} v_{+a}+\sigma_{+ac}v^c_{+}\right)u_{+b}
+\frac{1}{3}\left({\D}_{c} v^{c}_{+}\right) h^{+}_{ab}+{\D}_{\<a} v_{+b\>}+{\D}_{[a}v_{+b]}\,.
\end{eqnarray}
Up to the leading order in $v^a_{+}$ , the inverse of equation \eqref{eq:relativevelocity} is given by
$
u^a_{+}=\gamma\left(\tilde{u}^a_{+}+\hat{v}^{a}_{+}\right) \approx \tilde{u}^a_{+} + \hat{v}^a_{+} \,,
\hat{v}^{a}_{+}=-v^a_{+}\,,
$
where $\hat{v}^{a}_{+}\tilde{u}_{+a}=0$, and $\hat{v}^{a}_{+}v_{+a}=v^a_{+}v_{+a}$. 
 At the leading order in $v^a_{+}$ these observable quantities in both frames are related according to~\cite{Maartens:1998xg}
\begin{eqnarray}
\tilde{\Theta}_{+} &\approx& \Theta_{+} + {\D}_a v^a_{+}\,,
~~~~~~~~~~~~~
\tilde{\sigma}_{+ab} \approx  \sigma_{+ab} + {\D}_{\<a} v_{+b\>}\,,
\\
\tilde{\omega}_{+ab} &\approx& \omega_{+ab} + {\D}_{[a} v_{+b]}\,,
\label{eq:vorticty_transformation}
~~~~~~~~~~
\tilde{A}^a_{+} \approx A^a_{+}+ \frac{1}{3}\Theta v^a_{+} + \dot{v}^a_{+}\,.
\end{eqnarray}
The transformation between the components of the  energy-momentum tensor between these frames is given by~\cite{Maartens:1998xg,2012reco.book.....E}
\begin{eqnarray}\label{eq:energyflux}
\tilde{q}^a_{+} &\approx& q^a_{+}-\left( P_{+} + \rho_{m+}\right) v^a_{+} \,,
\qquad
\tilde{\rho}_{m+} \approx \rho_{m+} \,,
\qquad
\tilde{P}_{+} \approx P_{+}\,, \qquad \tilde{\pi}^{ab}_{+} \approx \pi^{ab}_{+} \,.
\end{eqnarray}
In General relativity, the fundamental 4-velocity is curl free, i.e ${\D}_{[b} A_{+c]} =0$, therefore from equation \eqref{eq:vorticity}, we have that $\omega_{+ab} = 0$, hence, $\tilde{\omega}_{+ab} = {\D}_{[a} v_{+b]}$. Also, on the homogenous background $\sigma_{+ab} =0$, therefore $\tilde{\sigma}_{+ab} = {\D}_{\<a} v_{+b\>}$.  Note that we can obtain a similar set of relations and equations for the collapsing region.

\section{Vorticity generation at the boundary}\label{sec:vorticit_generation}

In this section, we describe how vorticity is generated at the boundary of the two fluid elements due to the viscosity or relative friction at the boundary. 
The gradient of the pressure, the gravitational potential and the rate of expansion in the immediate neighbourhood of the boundary play a very important role. Note that the presence of vorticity invalidates the focusing theorem argument.

\subsection{Continuity and Euler equations in a dust dominated oriented universe}

Using the Bianchi identity $ \nabla_{[a} R_{bc]d}{}{}{}^c = 0$ and contracting it twice leads to $\nabla^a G_{ab} = 0$. Taking the divergence of equation \eqref{eq:GR} leads to the conservation equation for the total energy-momentum tensor, i.e. $\nabla_bT^{ab}_{\rm{tot}}=0$.
%Note that, in the rest frame of the fluid $\tilde{q}^a  = 0$ and $\tilde{A}^a=0$, 
The time and spatial components of  $\nabla_bT^{ab}_{\rm{tot}}=0$  at leading order in relative velocity is given by
\begin{eqnarray} \label{eq:matter_conservation}
 \dot{{\rho}}_{m+}  &+&\left({\rho}_{m+}+{P}_{+}\right)\left[ {\Theta}_{+}+ {\rm{Div}} v_{+} \right] =0\,,
 \\
 {\dot{v}}^{a}_{+} &+& \frac{1}{3} \Theta_{+} v^a_{+} + A^{a}_{+}
  + \frac{\dot{P}_{+}}{\left({\rho}_{+}+{P}_{+}\right)} v^a_{+} + \frac{{\D}^a P_{+}}{\left({\rho}_{+}+{P}_{+}\right)} 
% \\ && 
 -2\eta_{+}  {\rm{Div}}\sigma^a_{+}
 = 0\,,\label{eq:Euler_equation}
\end{eqnarray}
where ${\rm{Div}} \sigma_{+a} = {\D}^b \sigma_{+ab}$ is the divergence of the rate of shear deformation tensor and $ {\D}^b$ is a spatial derivative on the corresponding hypersurface. Again, a similar equation exists for the collapsing region. We split the boundary stress-energy tensor into collapsing and expanding parts (see equation \eqref{eq:shell_crossing_EMT} for details). The motivation for this approach is that there exists physical processes such as diffusion that transfer of energy/information from the boundary to the bulk. This is slightly different from cases where the conservation of the boundary stress-energy tensor is treatment as a separate unit \cite{Barrabes:1991ng}. The approach we discuss here is common in the field of hydrodynamics~\cite{Morton1984}
Equation \eqref{eq:matter_conservation} is the continuity equation or mater conservation equation and equation \eqref{eq:Euler_equation} is the Euler equation.
The $\dot{P}_{+}$ comes from the time derivative of equation \eqref{eq:energyflux} since $\tilde{q}^a$ vanishes in the rest-frame of the fluid
\begin{eqnarray}
\dot{q}^a_{+} =-\left( P_{+}+ \rho_{m+}\right) \dot{v}^a_{+} - \left( \dot{P}_{+} + \dot{\rho}_{m+}\right) v^a_{+}\,.
\end{eqnarray}
The Euler equation agrees with the results obtained in \cite{Maartens:1998xg} in the limit of vanishing viscosity. 
In addition, we have also set terms such as $ A^b_{+}\sigma_{+b}{}^a$, $\sigma_{+bc}\sigma^{bc}_{+} v^a_{+}$, to zero since they will only contribute at high order.  Going beyond the linear approximation in $v^a$  will be straight-forward.

Acting on equation \eqref{eq:Euler_equation} with a spatial derivative operator and taking the antisymmetric part gives the vorticity propagation equation
\begin{eqnarray}\label{eq:vorticity_equation}
 \dot{\tilde{\omega}}^{ab}_{+}+ \frac{1}{3} \Theta_{+} \tilde{\omega}^{ab}_{+} + \frac{\dot{P}_{+}}{\left({\rho}_{m+}+{P}_{+}\right)}\tilde{ \omega}^{ab}_{+}=2\eta_{+} {\D}^{[a}{\rm{Div}} \sigma^{b]}_{+} \,.
\end{eqnarray}
%AginNote that ${\D}^{[a}A^{b]} =0$ in  general relativity. 
The boundary pressure, $P_{+}$ acts like pressure associated with a Barotropic fluid $(P_{+} \propto \rho_{+m} )$. This can be seen from the definition of boundary pressure given in equation \eqref{eq:pressure_def}: $P_{+}=  \frac{1} {2\pi}\frac{1}{3}\left[\Theta_{+}\right] \propto \rho_{m+}$ ($\Theta_{+}$ is related to the matter density field through the continuity equation).  Thus, it is consistent to neglect its contribution to vorticity from terms such as 
$ {{\D}^{[a} P_{+}{\D}^{b]}\rho_{+m}}/{\left({\rho}_{m+}+{P}_{+}\right)^2} \rightarrow 0\,.$
Similarly,  we can use equation \eqref{eq:matter_conservation} to relate $\dot{P}_{+}$ to the square of  sound speed, 
$ {\partial P_{+}}/{\partial \rho_{+m}} = c_{s+}^2$:
 \begin{eqnarray}
\dot{P}_{+} = \frac{\partial P_{+}}{ \partial \rho_{m+}} \frac{\partial \rho_{m+}}{\partial t}  = c_{+s}^2 \dot{\rho}_{+} =  -c_{+s}^2 \left({\rho}_{m+}+{P}_{+}\right){\Theta_{+}} \longrightarrow  \frac{\dot{P}_{+}}{\left({\rho}_{m+}+{P}_{+}\right)} \tilde{\omega}^{ab}_{+} \approx - c_{+s}^2\bar{\Theta}_{+} \tilde{\omega}^{ab}_{+}\,.
\end{eqnarray}
However, $[[\Theta]]$ vanishes at the boundary, therefore, we set the contribution of $\dot{P}_{+}$ to zero.  
Using  the Ricci identity for the matter 4-vector: $2\nabla_{[c}\nabla_{b]} \tilde{u}_{+a} = R^d{}_{abc} \tilde{u}_{+d}$,
we show that $\eta_{+} {\D}^{[a}{\rm{Div}} \sigma^{b]}_{+} $ leads to vorticity diffusion term. 
This can be seen by acting on the Ricci identity with $\varepsilon_{abc}$  \cite{Ellisbook:2012}
\begin{eqnarray}\label{eq:constraint_eqn}
{\D}^{a} \tilde{\omega}^b_{+} + \frac{1}{3} \varepsilon^{abc}_{+}{\D}_{c}\tilde{\Theta}_{+} - \frac{1}{2} \varepsilon^{abc}_{+}{\D}^{d}\tilde{\sigma}_{+cd} =0 \,.
\end{eqnarray}
Taking the spatial derivative of equation \eqref{eq:constraint_eqn} and extracting the anti-symmetric part gives $\varepsilon_{abc} {\D}^{b}{\rm{Div}} \sigma^{c}_{+} = 2{\D}_{b} {\D}^b \tilde{\omega}_{+a}$. Putting all these back into equation \eqref{eq:vorticity_equation} gives
the propagation equation for the vorticity 
\begin{eqnarray}\label{eq:vorticity_eqn}
   \dot{\tilde{\omega}}_{+a}+\frac{1}{3}\tilde{\Theta}_{+}\tilde{\omega}_{+a}  =    %c_{+s}^2\tilde{\Theta}_{+} \tilde{\omega}_{+a} +
  \eta_{+} {\D}_{b} {\D}^b \tilde{\omega}_{+a}  \,.
\end{eqnarray}
where $\tilde{\omega}_{+a} = \varepsilon_{+abc}\tilde{ \omega}^{[bc]}_{+}/2$.  The second term on the RHS is a vorticity diffusion term. %This is due to the viscosity generated at the boundary between two fluid elements.  
%The first on the RHS is the acoustic dispersion term due to sound travelling at different speeds through different fluid elements in the neighbourhood of the boundary.  
The two terms on the LHS are derivative terms, this becomes clear when the Lie derivative operator is used to bring them together $ \mathcal{L}_{\tilde{u}} \tilde{\omega}_{+a} =    \dot{\tilde{\omega}}_{+a}+{\Theta}_{+}\tilde{\omega}_{+a} /3$.

\subsection{Vorticity generation  and the line of sight}\label{scalar_circulation}

Vorticity is one of the most important physical quantities in fluid dynamics. It is the most crucial observable for weather forecasting in the local environment~\cite{wang1993simple}. Vorticity gives a microscopic measure of the rotation at every point in the fluid flow.  As mentioned earlier, there is no true source term for vorticity in the vorticity propagation equation for the fluid (i.e equation \eqref{eq:vorticity_eqn}).  
The vorticity generation mechanism we describe here is built on the boundary layer theory developed by Morten (1984)~\cite{Morton1984} for incompressible fluids. 
Morton showed that the source for all vorticity in the fluid flow emanates from the boundary layers and
the rate at which vorticity enters the fluid from the boundary is determined by the viscous diffusion term.  
The rate of viscous diffusion is determined by the conditions imposed on circulation at the boundary layer. 

We discussed in sub-section \ref{sec:fluid_element} how vorticity is generated at the boundary between two different fluid elements in Cosmological context. {The Morten boundary layer theory fits naturally into our problem  because $\Theta$ vanishes at the boundary, therefore, the fluids act much like an incompressible fluid in the neighbourhood of the boundary.}   Our target here is to use this idea to show how vorticity is generated at the boundary between the expanding and collapsing regions of the universe.

Firstly, we would like to show how the radial component of the vorticity is generated. This component is very crucial for observational purposes because, for an observer, the radial component of vorticity could be compared to the  Kaiser redshift distortion term~\cite{Kaiser:1987qv}.  
To accomplish this, we decompose every observable into $1+1+2$ irreducible units. The first '1' denotes the time direction., the second '1' denotes the radial direction while the '2' denotes a closed 2-surface or the screen space.
For the radial direction, we use radial unit vector~$n^a$ with the following normalisation  $n^a_{+}n_{+a}=1$. $n^a_{+}$ is orthogonal to $u^a_{+}$:$~u^a_{+}n_{+a}=0$. The metric on the 2-surface is given by 
\begin{eqnarray}
N_{+a}{}^{b}\equiv h_{+a}{}^{b}-n_{+a}n^b_{+}=g_{+a}{}^{b}+u_{+a}u^b_{+}-n_{+a}n^b_{+},
\end{eqnarray}
$N_{+a}{}^{b}$  orthogonal to $n^a_{+}$ and $u^a_{+}$: $n^a_{+}N_{+ab}=0=u^a_{+}N_{+ab}$. It is used to project tensors onto 2-surfaces
 ($N_{+a}{}^{a}=2$). The spatial derivative of $n^a_{+}$ can be decomposed into  irreducible form~\cite{Clarkson:2007yp}:
\begin{eqnarray}\label{eq:fulldecompositionofDanb}
{\D}_a n_{+b}=n_{+a}\beta_{+b}+\frac{1}{2}\phi_{+} N_{+ab}+\xi_{+}\varepsilon_{+ab}+\zeta_{+ab},
\end{eqnarray}
where $\phi_{+}$ denotes the expansion/contraction of the 2-surface, $\zeta_{+ab}$  is the  shear distortion of the 2-surface, $\xi_{+}$ denotes the twisting of the 2-surface and $\hat{n}^a_{+}=\beta^b_{+}$  is the acceleration of $n^a_{+}$. The propagation equations for these terms are given in \cite{Clarkson:2007yp}.  We do not need their explicit form for the discussion that will follow. 
Furthermore, any 3-vector can be decomposed into radial components and tangential components. For example, we decompose the vorticity further as
\begin{eqnarray}\label{eq:omega_decomp}
\tilde{\omega}^a_{+}=\tilde{\omega}_{+\parallel} n^a_{+}+\tilde{\omega}^{a}_{+\bot},~~~{\rm{where}}~~~\tilde{\omega}_{+\parallel} 
\equiv \tilde{\omega}_{+a} n^a_{+},~~~\mbox{and}~~~\tilde{\omega}_{+\bot}^{a}\equiv
N^{ab}_{+}\tilde{\omega}_{+b} \,.
\end{eqnarray}
The acceleration vector $A^{a}$ is decomposed accordingly. 
The shear tensor on the spatial hypersurface is decomposed as 
\begin{eqnarray}
\sigma_{+ab}=\sigma_{+\parallel\parallel}\left(n_{+a}n_{+b}-\frac{1}{2}N_{+ab}\right)+2\sigma_{+(\bot\parallel a}n_{+b)}+\sigma_{+\bot\bot{ab}},
\label{tensor-decomp}
\end{eqnarray}
where we have introduced tensors that transforms as scalar, vector and tensor on the 2-surface
\begin{eqnarray}
\sigma_{+\parallel\parallel}&\equiv &n^a_{+}n^b_{+}\sigma_{+ab}=-N^{ab}_{+}\sigma_{+ab},~~~~~
\sigma_{+\bot \parallel a}\equiv N_{+a}^{~b}n^c_{+}\sigma_{+bc}\,,
\\ 
\sigma_{+\bot\bot{ab}}&\equiv &
\left({N_{+(a}^{~~c}N_{+b)}^{~~d}-\frac{1}{2}N_{+ab}N^{cd}_{+}}\right)\sigma_{+cd}\,.
\end{eqnarray}

\subsubsection{ Rate of change of scalar circulation}

\begin{figure}[h]
\includegraphics[width=100mm,height=70mm] {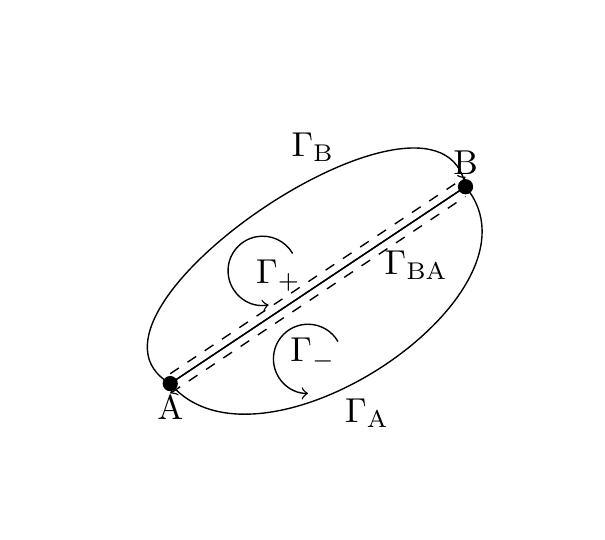}
\caption{{
Consider a projected sheet of two fluids in the region $\mathcal{S}={-,+}$ within the sheet $A_\mathcal{S}$. The sheet is  bounded by two curves  parallel to the interface between the collapsing and expanding regions. The circulation is calculated in the limit where the two curves approach the interface between the two regions/sheets. }}
%\red{I need to re-draw this}
\label{fig:patch_universe}
\end{figure}

The vorticity is related to the circulation through the Stokes theorem
\begin{eqnarray}\label{eq:scalar_circulation}
\Gamma = \oint_{\partial S} v_{a} {\d} {\rm{l}}^a =\int_{S} \tilde{\omega}_{a} {\d} {\rm{S}}^a = \int_{S} n^a {\omega} {\d} S= \int_{S} \tilde{\omega}_{\parallel} {\d} S=\oint_{\partial S} v_{a}  t^a {\d} {\rm{s}}\,,
\end{eqnarray}
where ${\d} {\rm{S}}^a$ is an oriented sheet orthogonal to the spacelike vector  $n^a$: ${\d} {\rm{S}}^a = n^a  {\d} S$. The surface $S$ is bounded by a closed contour $ {\d} {{l}}^a$ and $t^a$ is a 2-vector on the sheet.  The circulation defined in equation \eqref{eq:scalar_circulation} is more appropriately known as scalar circulation since it probes only the component of the vorticity parallel to $n^a$.  The total circulation within the universe with two different regions  as shown in figure \ref{fig:patch_universe} is given by
\begin{eqnarray}\label{eq:total_circulation}
\Gamma  %= \oint_{C} v_{\rm{Tot} a}\, {\d} {\rm{s}}^a 
= \Gamma_{-} + \Gamma_{+} + \int_{A}^{B} \gamma {\d} s \,,
\end{eqnarray}
where $\Gamma_{-}  $ and $\Gamma_{+}  $ is the circulation in $A_{-}$ and $A_{+}$ respectively.
 $ \int_{A}^{B} \gamma \,{\d} s$ is the circulation contained in the interface region and  $\gamma = \doublesqbraket{v_a {\hat{t}}^a} =\left[{v_{+}}_a  -{ v_{-}}_a \right]{\hat{t}}^a $ is the density of circulation contained within the boundary region due to a junction in the velocity vector. $A$ and $B$ are the limits of the integrals.  
 The full spacetime decomposition of the covariant derivative of $u^a$ in $1+3$ is given in equation \eqref{eq:covdu} and in $1+1+2$ is given by
\begin{eqnarray}
\nabla_au_b&=& -u_a\left({A}_{\parallel} n_b+\mathcal{A}_{\bot b}\right)+ n_a n_b\left(\frac{1}{3}\Theta+\sigma_{\parallel\parallel} \right)
+n_a \left(\sigma_{\parallel \bot b}+\varepsilon_{bc}\omega^c_{\bot}\right)
+\left(\sigma_{\parallel \bot a}-\varepsilon_{ac}\omega^c_{\bot}\right)n_b
\\ \nonumber &&
+N_{ab} \left(\frac{1}{3}\Theta-\frac{1}{2}\sigma_{\parallel\parallel}\right)+
\omega_{\parallel}\varepsilon_{ab}+\sigma_{\bot\bot ab}\,.
\label{eq:nabu112}
\end{eqnarray}
The time derivative of circulation in each time slice can be performed with the help of the fundamental theorem of calculus: 
 \begin{eqnarray}\label{eq:integral_formula}
 { \left.{\frac {d}{dt}}\right|_{t=0}\int _{ V}A =\int _{V }{\mathcal {L}}_{u }A } \,,
 \end{eqnarray}
 where $A = A_{a} \d x^a$ is a 1-form  and its Lie derivative is given by $\mathcal{L}_{u} \left(A_{a}  {\d} {\rm{x}}^{a}\right) = \left(u^b{\nabla}_{b}  A_{a} +   A_{b} {\nabla}_{a} u^b \right) {\d} x^ a$. Note that $u^a$ is the threading 4-vector.   We compute the time rate of change of circulation along the threading 4-vector% (it is equivalent to the fluid vector at linear order) 
 \begin{eqnarray}
{\frac {d}{dt}}\bigg|_{t=0} \Gamma &=&{\frac{d}{dt}}\bigg|_{t=0}\int_{A_{-}}  \tilde{\omega}_{-a} {\d} {\rm{S}}^{a}_{-}+  {\frac {d}{dt}}\bigg|_{t=0}\int_{A_{+}}  \tilde{\omega}_{+a}{\d} {\rm{S}}^{a}_{+} + {\frac {d}{dt}}\bigg|_{t=0}\int_{ A}^{B}  \gamma \,{\d} s\,,
\\
&=& \int_{A_{-}} {\mathcal {L}}_{u } \tilde{\omega}_{-a} {\d} {\rm{S}}^{a}_{-}+ \int_{A_{+}} {\mathcal {L}}_{u }\tilde{\omega}_{+a}{\d} {\rm{S}}^{a}_{+} 
+   {\frac {d}{dt}}\bigg|_{t=0} \int_{ A}^{B}  \gamma \,{\d} s \,,
\label{eq:Lie_inside-integral}
 \end{eqnarray}
 where the Lie derivative of vorticity is given in equation \eqref{eq:vorticity_eqn} and  ${\d} {\rm{S}}^{a} = n^a {\d} S$.
 To perform the integral over a closed 2-sphere, we need to further decompose $\tilde{\omega}^a$ into components living on a closed 2-sphere with one component parallel to $n^a$.
 % and the other orthogonal to $n^a$ (see equation \eqref{eq:omega_decomp}).
 Further decomposition of the spatial Laplacian of  $\tilde{\omega}^a$ is given by
 \begin{eqnarray}
{\D}_{b} {\D}^b \tilde{\omega}_{a} &=& 
n_{a} \bigg[ {\D}^2_{\parallel} \tilde{\omega}_{\parallel}
+ {\D}_{\parallel}\tilde{\omega}_{\parallel}
-\frac{1}{2} \phi^2 \tilde{\omega}_{\parallel} + \nabla_{\bot b} \nabla_{\bot}^b \tilde{\omega}_{\parallel}\bigg]
\\ \nonumber &&
+{\D}^2_{\parallel} \tilde{\omega}_{\bot a}+ \phi \left[{\D}_{\parallel} \tilde{\omega}_{\bot a} + \nabla_{\bot a} \tilde{\omega}_{\parallel}\right] - \frac{1}{4}\phi^2 \tilde{\omega}_{\bot a}  + \nabla_{\bot b} \nabla_{\bot}^{b} \tilde{\omega}_{\bot a}\,.
\end{eqnarray}
Note that $\phi$ is just an inverse of the fixed radial distance. 
We require that both the first and second radial derivatives of $\tilde{\omega}_{a}$ on closed 2-sphere vanish, thereby leading to a much-simplified expression
 \begin{eqnarray}
{\D}_{b} {\D}^b \tilde{\omega}_{a} &=& 
n_{a} \bigg[
-\frac{1}{2} \phi^2 \tilde{\omega}_{\parallel} + \nabla_{\bot b} \nabla_{\bot}^b \tilde{\omega}_{\parallel}\bigg]
+ \phi  \nabla_{\bot a} \tilde{\omega}_{\parallel} - \frac{1}{4}\phi^2 \tilde{\omega}_{\bot a}  + \nabla_{\bot b} \nabla_{\bot}^{b} \tilde{\omega}_{\bot a} \,.
\end{eqnarray}
Now we are in a position to evaluate the integrals in equation \eqref{eq:Lie_inside-integral}.
Note that the total circulation around the whole loop is the sum of the circulations around the two loops:
\begin{eqnarray}\label{eq:loops}
\Gamma_{-}=\Gamma_{A}+\Gamma_{AB} \qquad {\rm{and}}\qquad  \Gamma_{+}=\Gamma_{A}+ \Gamma_{BA} \,,
\end{eqnarray}
where the circulation around $\Gamma_{-}$  is the sum of an integral along $\Gamma_{A}$ and along $\Gamma_{BA}$. Similarly, the circulation around $\Gamma_{+}$ is the sum of two parts, one along $\Gamma_{B}$ and the other along $\Gamma_{AB}$. The integral along $\Gamma_{AB}$ differs from  $\Gamma_{BA}$ by an opposite sign. This is because the direction of travel is opposite.
The radial component of the vorticity propagation equation becomes
 \begin{eqnarray}\label{eq:radial_vorticity}
  n^a \mathcal{L}_{u} \tilde{\omega}_a  &=&   \eta \nabla_{\bot b} \nabla_{\bot}^b \tilde{\omega}_{\parallel} \,.
  %%+ \frac{1}{2} \eta \phi^2 \tilde{\omega}_{\parallel}  %+ \left[c_{s}^2 {\Theta}  \right] \tilde{\omega}_{\parallel} \,.
 \end{eqnarray}
 {Note that $\phi$ vanishes at the boundary since $\Theta$ vanishes. }
Putting equation \eqref{eq:radial_vorticity} in equation \eqref{eq:Lie_inside-integral} while remembering  to use equation \eqref{eq:loops}, we can perform the integration using equation \eqref{eq:scalar_circulation} and Divergence theorem gives~\cite{terringtonhouriganthompson:2022}:
\begin{eqnarray} 
{\frac {d}{dt}}\bigg|_{t=0} \Gamma &=&\oint_{\partial S_{-}}  \eta_{-} t^a \nabla_{\bot a} \tilde{\omega}_{-\parallel} {\d} {{s}} 
%+  \oint_{\partial S_{-}} \left[c_{s-}^2 {\Theta}_{-} + \frac{1}{2} \eta_{-}\phi^2_{-} \right] v_{-a}  t^a {\d} {{s}} 
+\oint_{\partial S_+}  \eta_{+} t^a \nabla_{\bot a} \tilde{\omega}_{+\parallel} {\d} {{s}} 
%\\  \nonumber  &&
%+  \oint_{\partial S_+} \left[c_{s+}^2 {\Theta}_{+} + \frac{1}{2} \eta_{+}\phi^2 \right] v_{+a}  t^a {\d} {{s}} 
+ \int_{ A}^{B}\left(\Sigma_{-} + \Sigma_{+}\right)\,{\d} s 
%+\int_{ A}^{B} \left(\lambda_{-} +\lambda_{+}\right)\,{\d} s  
+{\frac {d}{dt}}\bigg|_{t=0}\int_{ A}^{B}  \gamma \,{\d} s \,,
\label{eq:introduc_vorticity_flux}
\end{eqnarray}
where $\Sigma_{-} = \eta_{-} t^a \nabla_{\bot a} \tilde{\omega}_{-\parallel}  $ and  $ \Sigma_{+} =  -\eta_{+} t^a \nabla_{\bot a} \tilde{\omega}_{+\parallel} $ is the diffusive vorticity flux of vorticity from the fluid in $-$ and $+$ respectively. 
%While $\lambda_{-} = -\left[c_{s-}^2 {\Theta}_{-} + \frac{1}{2} \eta_{-}\phi^2_{-} \right] {v_{-}}_{a}t^a$ and  $\lambda_{+} =  \left[c_{s+}^2 {\Theta}_{+} + \frac{1}{2} \eta_{+}\phi^2_{+} \right] {v_{+}}_{a}t^a$ are acoustic vorticity flux term due to the time rate of change of pressure and projection effects. 
 %
 To evaluate the rate of change of the  $\gamma$, we have to, first of all, separate the terms 
  \begin{eqnarray}
 {\frac {d}{dt}}\bigg|_{t=0}\int_{ A}^{B}  \gamma \,{\d} s  = {\frac {d}{dt}}\bigg|_{t=0}\int_{ A}^{B}  {v_{+}}_{a} \,{\d} {\rm{s}}^a  -  {\frac {d}{dt}}\bigg|_{t=0}\int_{ A}^{B}  {v_{-}}_{a} \,{\d} {\rm{s}}^a\,.
 \end{eqnarray}
We can make use of equation \eqref{eq:integral_formula} to evaluate the time rate of change of the integral. Note that to do this consistently we will need a $1+1+1+1$  decomposition equivalent of equation \eqref{eq:nabu112}. However, since we are considering only the linear order approximation, we can proceed as in equation \eqref{eq:Lie_inside-integral}
 \begin{eqnarray}
  {\frac {d}{dt}}\bigg|_{t=0}\int_{ A}^{B}  \gamma \,{\d} s  &=&\int_{ A}^{B}    {\mathcal {L}}_{u }v_{+}^{a} \,{\d} {\rm{s}}^a- \int_{ A}^{B}    {\mathcal {L}}_{u }v_{-}^{a} \,{\d} {\rm{s}}^a
 =\int_{ A}^{B}    t_a{\mathcal {L}}_{u }v_{+}^{a} \,{\d} {{s}} - \int_{ A}^{B}   t_a {\mathcal {L}}_{u }v_{-}^{a} \,{\d} {{s}}\,.
 \label{eq:velocity_jump}
 \end{eqnarray}
 The circulation within the interface is generated by the relative acceleration of fluid elements on both sides of the boundary. 
 The Lie derivative of closed 2-sphere projected relative velocity (i.e $ {\mathcal {L}}_{u }{v_{+}}_{a} $),  is given by the  momentum conservation equation 
 \begin{eqnarray}\label{eq:2dprejected_momentum}
 {\mathcal {L}}_{u }{v_{\bot}}^{a}    &=& -A^{a}_{\bot}
  - \frac{{\D}^a_{\bot} P}{\left({\rho}+{P}\right)} 
 +\eta\frac{2}{3} {\nabla}^{a}_{\bot} \tilde{\Theta} %+c_{s}^2 {\Theta} v^a_{\bot} 
 - \varepsilon_{ab}\eta %\frac{1}{2} \phi\tilde{\omega}_{\bot}^b 
 \nabla_{\bot}^b \tilde{\omega}_{\parallel} \,,
\end{eqnarray}
where $ {\mathcal {L}}_{u }v_{\bot}^{a}  = {\dot{v}}^{a}_{\bot}  + \Theta v^a_{\bot}/3 $ is a coordinate independent acceleration term.  Note that at leading order and assuming the FLRW background $\dot{n}^a\approx 0$.  %Equation \eqref{eq:2dprejected_momentum} is a statement of momentum conservation. 
Contracting equation \eqref{eq:2dprejected_momentum} with a 2-vector $t^a$ gives 
 \begin{eqnarray}\label{eq:2dprejected_momentum2}
t_{a} {\mathcal {L}}_{u }{v_{\bot}}^{a}    &=& -t_{a}A^{a}_{\bot}
  - \frac{t_{a}{\nabla}^a_{\bot} P}{\left({\rho}+{P}\right)} 
 +\eta  \frac{2}{3}t_{a} {\nabla}^{a}_{\bot} \tilde{\Theta} %+\frac{1}{2} \eta\phi t_{a} \tilde{\omega}_{\bot}^a %+c_{s}^2 {\Theta} t_{a}v^a_{\bot}
  -\eta t_{a}\nabla_{\bot}^a \tilde{\omega}_{\parallel}\,,
\end{eqnarray}
where $ \varepsilon_{ab}\nabla_{\bot }^b\tilde{ \omega}_{\parallel}\equiv  \nabla_{\bot a}\tilde{\omega}_{\parallel}  $ and at leading order $A_{\bot a} = {\nabla}_{\bot a}\Phi$. We introduce another directional directive $t^a\nabla_{\bot a} = {\partial }/{\partial s}$ to improve clarity.
%Furthermore, $t_{a} \tilde{\omega}_{\bot}^a$ can be expressed in terms of the projected relative velocity as 
%\begin{eqnarray}\label{eq:omegaperpv}
%t^d\tilde{\omega}_{\bot d}= t^dN_{d}{}^{a} \tilde{\omega}_{a} = \frac{1}{2}\varepsilon_{abc} t^dN_{d}{}^{a}{\D}^b v^c =  \phi v_{\bot a} t^a - 2 t^a \nabla_{\bot a} v_{\parallel}  =- 2 t^a \nabla_{\bot a} v_{\parallel} \,.
%\end{eqnarray}
%Plugging equation \eqref{eq:omegaperpv} into equation \eqref{eq:%2dprejected_momentum} gives
%\begin{eqnarray}
%t_{a} {\mathcal {L}}_{u }{v_{\bot}}^{a}    &=& -t_{a}A^{a}_{\bot}
%  - \frac{t_{a}{\nabla}^a_{\bot} P}{\left({\rho}+{P}\right)} 
% +\frac{2}{3}\eta t_{a} {\nabla}^{a}_{\bot} \tilde{\Theta} 
 %+\phi \eta t^a \nabla_{\bot a} v_{\parallel} +  \frac{1}{2}\phi^2 \eta t_{a}v^a_{\bot}
% -\eta t_{a}\nabla_{\bot}^a \tilde{\omega}_{\parallel}\,.
%\label{eq:massage_momentum}
%\end{eqnarray}
Identifying the last term in equation \eqref{eq:2dprejected_momentum2} with the diffusive flux  introduced in equation \eqref{eq:introduc_vorticity_flux} and  making both terms the subject of the expression leads to
\begin{eqnarray}
\Sigma_{-}  &=&- t_{a} {\mathcal {L}}_{u }{v_{-}}_{\bot}^{a} +t_{a}{A_{-}}^{a}_{\bot}
  + \frac{t_{a}{\nabla}^a_{\bot} P_{-}}{\left({\rho_{m-}}+{P_{-}}\right)} 
 - \frac{2}{3}{\eta_{-}}t_{a} {\nabla}^{a}_{\bot} \tilde{\Theta}_{-} 
 %-\phi_{-} {\eta_{-}}t^a \nabla_{\bot a} {v_{-}}_{\parallel}
  \,,
 \label{eq:sigma1lambda1}
 \\
 \Sigma_{+}  &=& t_{a} {\mathcal {L}}_{u }{v_{+}}_{\bot}^{a} -t_{a}{A_{+}}^{a}_{\bot}
  -\frac{t_{a}{\nabla}^a_{\bot} P_{+}}{\left({\rho_{m+}}+{P_{+}}\right)} 
 + \frac{2}{3}t_{a} {\eta_{+}}{\nabla}^{a}_{\bot} \tilde{\Theta}_{+}
 %+{\phi_{+}} {\eta_{+}}t^a \nabla_{\bot a} {v_{+}}_{\parallel} 
 \,.
  \label{eq:sigma2lambda2}
\end{eqnarray}
%\blue{check the signs}
Combing equations \eqref{eq:sigma1lambda1} and \eqref{eq:sigma2lambda2} give
\begin{eqnarray}\label{eq:viscous_accen}
\left(\Sigma_{-} + \Sigma_{+}\right) %+ \left(\lambda_{-} +\lambda_{+}\right) 
= - t_{a} \doublesqbraket{{\mathcal {L}}_{u }{v}_{\bot}^{a} }
-\doublesqbraket{ \frac{\partial \Phi }{\partial s} }
- \doublesqbraket{\frac{\partial }{\partial s}  \frac{P}{\left({\rho_{m}}+{P}\right)} } + \frac{2}{3}\doublesqbraket{{\eta} \frac{\partial \tilde{\Theta} }{\partial s}} 
%+ \doublesqbraket{  {\eta}{\phi}\frac{\partial v_{\parallel} }{\partial s}}
\,,
\end{eqnarray}
where the boundary vorticity flux is related to the boundary acceleration through the momentum equation.
This provides a source for the vorticity propagation equation \eqref{eq:radial_vorticity}. We can write the jump in the viscous  acceleration  of  fluid  elements  on  each  side  of the closed 2-sphere as 
\begin{eqnarray}
t_{a} \doublesqbraket{{\mathcal {L}}_{u }{v}_{\bot}^{a} }= -\left(\Sigma_{-} + \Sigma_{+}\right) %- \left(\lambda_{-} +\lambda_{+}\right)  
-\doublesqbraket{ \frac{\partial \Phi }{\partial s} }
- \doublesqbraket{\frac{\partial }{\partial s}  \frac{P}{\left({\rho_{m}}+{P}\right)} } + \frac{2}{3}\doublesqbraket{{\eta} \frac{\partial \tilde{\Theta} }{\partial s}} 
%+ \doublesqbraket{ {\phi} {\eta}\frac{\partial v_{\parallel} }{\partial s}} 
\,.
\label{eq:vis_accen}
\end{eqnarray}
The jump in viscous acceleration is sourced by the angular gradients in the gravitational potential, pressure and expansion.
Plugging equation \eqref{eq:vis_accen} into equation \eqref{eq:velocity_jump}  and performing the arc-length integration gives
\begin{eqnarray}\nonumber
  {\frac {d}{dt}}\bigg|_{t=0}\int_{ A}^{B}  \gamma \,{\d} s  &=&-\int_{ A}^{B} \left(\Sigma_{-} + \Sigma_{+}\right)  {\d} s 
  % - \int_{ A}^{B}  \left(\lambda_{-} +\lambda_{+}\right)  {\d} s
   - \left( \doublesqbraket{ \Phi_{B} }- \doublesqbraket{\Phi_{A}}\right) -  \doublesqbraket{\frac{1}{(1+w)} \frac{P}{\rho_{m}} \bigg|_{B}} 
    \\  &&
     +\doublesqbraket{\frac{1}{(1+w)} \frac{P}{\rho_{m}} \bigg|_{A} }
  + \frac{2}{3} \doublesqbraket{ {\eta}\tilde{\Theta}_{B} }- \frac{2}{3}\doublesqbraket{{\eta}\tilde{\Theta_{A}}}
  \,.
  %+  \doublesqbraket{ {\eta}{\phi}v_{\parallel B} }- \doublesqbraket{{\eta}{\phi} v_{\parallel A}}
\end{eqnarray}
%$\frac{1}{(1+w)} \frac{P}{\rho} $
The circulation within the boundary is generated by the relative acceleration of fluid elements on both sides of the boundary which is sourced by the differences in gravitational potential, the sum of pressure and matter density and relative expansion. The last two terms on the second line  vanishes at the boundary since $\Theta =0$. Neglecting these terms, we recover the standard result in boundary layer vorticity generation~\cite{terringtonhouriganthompson:2022,bronsthompsonewekehourigan:2014}
\begin{eqnarray}\label{eq:circulation_condtions}
  {\frac {d}{dt}}\bigg|_{t=0}\int_{ A}^{B}  \gamma \,{\d} s  &=&-\int_{ A}^{B} \left(\Sigma_{-} + \Sigma_{+}\right)  {\d} s  
  - \left( \doublesqbraket{ \Phi_{B} }- \doublesqbraket{\Phi_{A}}\right) -  \doublesqbraket{\frac{1}{(1+w)} \frac{P}{\rho_{m}} \bigg|_{B}} 
     +\doublesqbraket{\frac{1}{(1+w)} \frac{P}{\rho_{m}} \bigg|_{A} }\,.
\end{eqnarray}
In this limit,  circulation within the interface is generated by the differences in gravitational potential and the ratio of pressure to the matter density. Although there is a  flux  of  vorticity  term in equation \eqref{eq:circulation_condtions}, it does not contribute to the net rate of circulation
\iffalse
\begin{eqnarray}\nonumber
{\frac {d}{dt}}\bigg|_{t=0} \Gamma &=&\oint_{\partial S_{-}}  \eta_{-} t^a \nabla_{\bot a} \tilde{\omega}_{-\parallel} {\d} {{s}} 
%+  \oint_{\partial S_{-}} \left[c_{-s}^2 {\Theta}_{-} + \frac{1}{2} \eta_{-}\phi^2 \right] v_{-a}  t^a {\d} {{s}} 
+\oint_{\partial S_{+}}  \eta_ {+}^a \nabla_{\bot a} \tilde{\omega}_{+\parallel} {\d} {{s}} 
%\\ \nonumber &&
   \\  &&
%+  \oint_{\partial S_{+}} \left[c_{s+}^2 {\Theta}_{+} + \frac{1}{2} \eta_{+}\phi^2 \right] v_{+a}  t^a {\d} {{s}} 
- \left( \doublesqbraket{ \Phi_{B} }- \doublesqbraket{\Phi_{A}}\right) -  \doublesqbraket{\frac{1}{(1+w)} \frac{P}{\rho} \bigg|_{B}} 
     +\doublesqbraket{\frac{1}{(1+w)} \frac{P}{\rho} \bigg|_{A} }\,.
  + \frac{2}{3} \doublesqbraket{ {\eta}\tilde{\Theta}_{B} }- \frac{2}{3}\doublesqbraket{{\eta}\tilde{\Theta_{A}}}
  %+  \doublesqbraket{ {\eta}{\phi}v_{\parallel B} }- \doublesqbraket{{\eta}{\phi} v_{\parallel A}}
  \,.
\end{eqnarray}
Again neglecting the general relativistic  correction appearing due to use of the Ricci identity  leads to 
\fi
\begin{eqnarray}\nonumber 
{\frac {d}{dt}}\bigg|_{t=0} \Gamma &=&\oint_{\partial S_{-}}  \eta_{-} t^a \nabla_{\bot a} \tilde{\omega}_{-\parallel} {\d} {{s}} 
%+  \oint_{\partial S_{-}} \left[c_{-s}^2 {\Theta}_{-} + \frac{1}{2} \eta_{-}\phi^2 \right] v_{-a}  t^a {\d} {{s}} 
+\oint_{\partial S_{+}}  \eta_{+} t^a \nabla_{\bot a} \tilde{\omega}_{+\parallel} {\d} {{s}} 
\\  &&
%+  \oint_{\partial S_{+}} \left[c_{+s}^2 {\Theta}_{+} + \frac{1}{2} \eta_{+}\phi^2_{+}\right] v_{+a}  t^a {\d} {{s}} 
- \left( \doublesqbraket{ \Phi_{B} }- \doublesqbraket{\Phi_{A}}\right) -  \doublesqbraket{\frac{1}{(1+w)} \frac{P}{\rho_{m}} \bigg|_{B}} 
%\\ \nonumber &&
     -\doublesqbraket{\frac{1}{(1+w)} \frac{P}{\rho_{m}} \bigg|_{A} }
 \,.
\end{eqnarray}
The flux diffusion term transport vorticity from the boundary to the interior of the fluid, they do not generate vorticity. The vorticity is generated by the difference in the gravitational and pressure across the boundary. 
Moreover, if one imposes the smoothness of the gravitational potential at the boundary $\doublesqbraket{ \Phi_{B} }=\doublesqbraket{\Phi_{A}}=0$, then the vorticity can only be generated by the differences in pressure across the interface, otherwise the circulation is globally conserved.
A mutual generation and annihilation of vorticity could still happen in the neighbourhood of the boundary. 

Finally, if we further impose no-slip condition  at the boundary, we find a simple expression for the vorticity flux across the interface
\begin{eqnarray}\label{eq:viscous_accen2}
\left(\Sigma_{-} + \Sigma_{+}\right) = 
-\doublesqbraket{ \frac{\partial \Phi }{\partial s} }
- \doublesqbraket{\frac{\partial }{\partial s}  \frac{P}{\left({\rho}+{P}\right)} } + \frac{2}{3}\doublesqbraket{{\eta} \frac{\partial \tilde{\Theta} }{\partial s}} \,.
\end{eqnarray}
Here the vorticity flux is sourced exclusively by the pressure, gravitational potential  and expansion gradients, this is in agreement with \cite{bronsthompsonewekehourigan:2014,terringtonhouriganthompson:2020}.  %This is one of our key results. 

\subsection{Vorticity generation at the boundary and the rate of change of vector circulation}

In this sub-section, we extend the treatment to the vorticity vector on the hypersurface. 
The basic building block of this is the curl theorem. It  relates vorticity at a given instant in time to the relative velocity~\cite{terringtonhouriganthompson:2022}
 \begin{eqnarray}\label{eq:3Dboundary_integration}
 {{\Gamma}_a} &=& \int_{V}{\tilde{ \omega}}_a \,\d V=\int_{\partial V}   \varepsilon_{abc}   {{v}^c} \,{\d \rm{S}}^b   = \int_{\partial V}   \varepsilon_{abc} {n}^b  {{v}^c} \,{\d} S\,,
 \end{eqnarray}
 where $ {{\Gamma}_a} $ is called a vector circulation, it gives the macroscopic picture of rotation within a given local region and $V$ is the volume of a given hypersurface, $\partial V$ denotes the boundary of the hypersurface, 
The total vector circulation  of the universe at a given time slice is given by the  sum of vorticity in the fluids in the two regions separated by a shell and the vorticity within the shell~\cite{terringtonhouriganthompson:2022}
\begin{eqnarray}
\Gamma_{a}  =  \int_{V_{-}}  \tilde{\omega}_{-a} \d V_{-} + \int_{V_{+}}  \tilde{\omega}_{+ a} {\d} V_{+} + \int_{ I}   \varepsilon_{abc} {n}^{b} \left( v_{+}^{c} -v_{-}^{c} \right) \,\d S \,.
\end{eqnarray}
Note that $n^a$ is pointing in the direction of increasing radial distance. %$ \theta_A$ and ${ \theta_B}$ are the angular position of the shells. 
 The time derivative of the  vector circulation is given by
\begin{eqnarray}\label{eq:time_deri_Gamma}
{\frac {d}{dt}}\bigg|_{t=0} \Gamma_a={\frac {d}{dt}}\bigg|_{t=0}  \int_{V_{-}}  \tilde{\omega}_{-a} \d V_{-} +{\frac {d}{dt}}\bigg|_{t=0}  \int_{V_{+}}  \tilde{\omega}_{+ a} \d V_{+}    + {\frac {d}{dt}}\bigg|_{t=0} \int_{ I}   \gamma_a  \,\d S \,,
\end{eqnarray}
where $\gamma_a = \varepsilon_{abc} {n}^{b} \left( {{v}^{c}_{+}} -{{v}^{c}_{-}} \right) $ is the density of circulation contained within the boundary region due to a possible jump in the velocity vector. 
Using equation \eqref{eq:integral_formula}, we evaluate the time derivative over the integral  and substituting  for the Lie derivative of $\tilde{\omega}_a$ using equation \eqref{eq:vorticity_equation} leads to 
\begin{eqnarray}
{\frac {d}{dt}}\bigg|_{t=0} \Gamma_a&=&   \int_{V}\left[ \eta  {\D}_{b} {\D}^b \tilde{\omega}_{a}  %+c_{s}^2 {\Theta}\tilde{ \omega}^{a}
\right]  \d V\,.
\end{eqnarray}
%\red{check the signs}
Performing the integration in equation \eqref{eq:time_deri_Gamma} while taking into account the possible discontinuity across the shell  gives
\begin{eqnarray}%\nonumber
{\frac {d}{dt}}\bigg|_{t=0} \Gamma_a
&=&    \int_{\partial V_{-}}\eta_{-} {n}_{b} {\D}^b \tilde{\omega}_{-a}  \d S_{-} 
%+\int_{\partial V_{-}}c_{s{-}}^2 {\Theta} \varepsilon_{abc}n^c  {v}^b_{-} \d S_{-} 
 + \int_{\partial V_{+}}\eta_{+} {n}_{b} {\D}^b \tilde{\omega}_{+a}  \d S_{+} 
 %+\int_{\partial V_{+}}c_{s{+}}^2 {\Theta} \varepsilon_{abc}n^c  {v}^b_{+} \d S_{+}
%\\  &&
+\int_{ I}  \left(\Sigma_{+a} +\Sigma_{{-}a}\right)   \d S 
%+\int_{ I} \left(\lambda_{+a} + \lambda_{{-}a}\right) \d S
%\\ \nonumber  &&
 + {\frac {d}{dt}}\bigg|_{t=0} \int_{I}   \gamma_a  \,\d S\,,
 \label{eq:vctor_cir_gamma}
\end{eqnarray}
where we have introduced  the viscous vorticity flux and the  acoustic flux in the neighbourhood of the shell 
\begin{eqnarray}\label{diffusive_flux_vectors}
\Sigma_{{-}a} &=& \eta_{{-}}n_{b}{\D}^b \tilde{\omega}_{{-}a}\,, \qquad\qquad
\Sigma_{{+}a} = -\eta_{+}n_{b}{\D}^b \tilde{\omega}_{{+}a}\,.
%\\
%\lambda_{{-}a} &=&-c_{s{-}}^2 {\Theta}_{+}  \varepsilon_{ab}  {v}^b_{-}\,, \qquad\qquad\quad
%\lambda_{{+}a} = c_{s{+}}^2 {\Theta}_{+}    \varepsilon_{ab}  {v}^b_{+}\,.
\label{advective_vectors}
\end{eqnarray}
 %where $ \varepsilon_{ab} = \varepsilon_{abc} n^c$. 
For the integrals containing the viscous terms in equation \eqref{eq:vctor_cir_gamma}, we made use of the Divergence theorem to perform the integration.% and for the integrals containing the relative expansion term, we made use of equation \eqref{eq:3Dboundary_integration}
 To evaluate the rate of change of the vector circulationl, we have to first of all separate  out the terms for clarity
 \begin{eqnarray}
 {\frac {d}{dt}}\bigg|_{t=0} \int_{ I}  \gamma_{a} \,{\d} S  &=& {\frac {d}{dt}}\bigg|_{t=0} \int_{ I}  \varepsilon_{abc} {n}^{c}{v_{2}}^{b} \,{\d} {S}  -  {\frac {d}{dt}}\bigg|_{t=0}\int_{ I}  \varepsilon_{abc} {n}^{c} {v_{{-}}}^{b} \,{\d} {S}\,,
 \\
  &=& \int_{ I}  \varepsilon_{ab}  ( \mathcal{L}_{u}{v_{2}}^{b}) \,{\d} {S} + \int_{ I}(\dot{ \varepsilon}_{ab} )  {v_{2}}^{b} \,{\d} {S} 
   - \bigg[\int_{ I}  \varepsilon_{ab} ( \mathcal{L}_{u}{v_{{-}}}^{b} )\,{\d} {S} +\int_{ I}   (\dot{\varepsilon}_{ab}) {v_{{-}}}^{b} \,{\d} {S} \bigg]\,.
 \end{eqnarray}
 The directional  derivative of ${ \varepsilon}_{ab} $  along $u^a$ is given by 
 \begin{eqnarray}
 \dot{ \varepsilon}_{ab}&=&-2u_{[a} \varepsilon_{b]c}A^c_{\bot}+2n_{[a} \varepsilon_{b]c}\dot{n}^c,
 \end{eqnarray}
 where $A^c_{\bot}$ is a closed 2-sphere projected acceleration vector.  The directional  derivative of $n^a$  along $u^a$ {is given by}\cite{Clarkson:2007yp}	
\begin{eqnarray}
\dot{ n}_{a}={A_{\parallel}} \dot{u}_a+\alpha_a~~~\mbox{where}~~~\alpha_a 	\equiv\dot n_{\bar a}
~~~\mbox{and}~~~\mathcal{A}=n^a\dot{u}_a.
\end{eqnarray}
These terms vanish on an FLRW background, similarly,  $ \dot{ \varepsilon}_{ab}$  and ${\sigma}^c{}_{[a}{ \varepsilon}_{b]c}$ vanish, hence the Lie derivative of ${\gamma}_{a}$ at linear order in relative velocity is given by
\begin{eqnarray}\label{eq:simplified_junction_condition}
 {\frac {d}{dt}}\bigg|_{t=0} \int_{ I}   \gamma_{a} \,{\d} S  &=& \int_{ I} \varepsilon_{ab}  ( \mathcal{L}_{u}{v_{2}}^{b}) \,{\d} {S}  - \int_{ I}  \varepsilon_{ab} ( \mathcal{L}_{u}{v_{{-}}}^{b} )\,{\d} {S} \,.
\end{eqnarray}
Using the momentum constraint equation, we find that the Lie derivative of $v^a$ is given by
 \begin{eqnarray}
 \varepsilon_{ab}  \mathcal{L}_{u} v^b =  \varepsilon_{ab} \left[ {\dot{v}}^{b}  +\frac{1}{3} \Theta v^b \right]&=& -\varepsilon_{ab} A^{b} 
  - \varepsilon_{ab}\frac{ {\D}^b P}{\left({\rho_{m}}+{P}\right)} 
 +\frac{2}{3}\eta   \varepsilon_{ab} {\D}^{b} \tilde{\Theta}-\eta n_{c}{\D}^c \tilde{\omega}^b 
 %+c_{s}^2 {\Theta} \varepsilon_{ab} v^a 
 \,.
 \label{eq:3Dmomentum_eqn}
 \end{eqnarray}
 Note that we can now define the projected angular derivatives and projected tensors
 \begin{eqnarray}
 \varepsilon_{ab} A^{b}  =  \varepsilon_{ab}A^{b}_{\bot} \equiv A_{\bot a} = \nabla_{\bot a} \Phi\,,  ~~~~~\varepsilon_{ab}{\D}^b P  \equiv  \nabla_{\bot a} P ~~~~~
 {\rm{and}}  ~~~~~~  \varepsilon_{ab} {\D}^{b} \tilde{\Theta}\equiv  \nabla_{\bot a}  \tilde{\Theta}\,.
 \end{eqnarray}
 The last two terms in equation \eqref{eq:3Dmomentum_eqn} can be identified with the terms defined in equations 
\eqref{diffusive_flux_vectors} and \eqref{advective_vectors}. Making these terms the subject of the formulae leads to 
\begin{eqnarray}
\Sigma_{{-}a}  &=& -  \mathcal{L}_{u} v_{{-}\bot a} -A_{{-}\bot a}
  - \frac{ {\nabla}_{\bot a} P_{{-}}}{\left({\rho_{m-}}+{P_{-}}\right)} 
 +\frac{2}{3}\eta_{{-}}  {\nabla}_{\bot a} \tilde{\Theta}_{-}\,,
 \\
 \Sigma_{{+}a}  &=&   \mathcal{L}_{u} v_{{+}\bot a} +A_{{+}\bot a}
  + \frac{ {\nabla}_{\bot a} P_{+}}{\left({\rho}_{m+}+{P}_{+}\right)} 
 -\frac{2}{3}\eta_{{+}}  {\nabla}_{\bot a} \tilde{\Theta}_{+}\,,
\end{eqnarray}
 where we take the following  approximation $ \varepsilon_{ab}  \mathcal{L}_{u} v^b =   \mathcal{L}_{u} v_{\bot a}$. Combing these two equations leads to the expression of the diffusive flux vector in terms of the acceleration term, angular pressure gradient, gravitational potential gradients and the gradient in the expansion rates
 \begin{eqnarray}\label{eq:vorticity_flux_and_advective}
 \left( \Sigma_{+a}+ \Sigma_{{-}a} \right) %+ \left( \lambda_{+a} +  \lambda_{{-}a}\right) 
 =  \doublesqbraket{  \mathcal{L}_{u} v_{\bot a}}
 + \doublesqbraket{A_{\bot a}} +  \doublesqbraket{ \frac{ {\nabla}_{\bot a} P}{\left({\rho_{m}}+{P}\right)} }  
 - \frac{2}{3}\doublesqbraket{  \eta  {\nabla}_{\bot a} \tilde{\Theta}}\,.
 \end{eqnarray}
Equation \eqref{eq:vorticity_flux_and_advective} indicates that the effective vorticity flux out of a given shell is
equal to the difference in acceleration of fluid elements on both sides of the shell. Imposing the no-slip Junction conditions for the geodesic (i.e equation \eqref{eq:four_velocity})
$
 \doublesqbraket{  \mathcal{L}_{u} v_{\bot a}} =  \mathcal{L}_{u}\doublesqbraket{   v_{\bot a}}  = \mathcal{L}_{u} \left[ v_{\bot +a} -v_{\bot {-}a}\right]= 0
$
 \begin{eqnarray}\label{eq:vorticity_flux_and_advective2}
 \left( \Sigma_{+a}+ \Sigma_{{-}a} \right) %+ \left( \lambda_{+a} +  \lambda_{{-}a}\right) 
 = 
  \doublesqbraket{A_{\bot a}} +  \doublesqbraket{ \frac{ {\nabla}_{\bot a} P}{\left({\rho_{m}}+{P}\right)} }  
 - \frac{2}{3}\doublesqbraket{  \eta  {\nabla}_{\bot a} \tilde{\Theta}}\,.
 \end{eqnarray}
 The gradients in pressure, gravitational field and expansion across the shell act as sources for vorticity in the universe. 
The momentum constraint equation at shell crossing is given by
 \begin{eqnarray}\label{eq:junction_momentum_eqn}
  \doublesqbraket{  \mathcal{L}_{u} v_{\bot a}} =  \left( \Sigma_{+a}+ \Sigma_{{-}a} \right) %+ \left( \lambda_{+a} +  \lambda_{{-}a}\right)
  - \doublesqbraket{\nabla_{\bot a} \Phi} -  \doublesqbraket{ \frac{ {\nabla}_{\bot a} P}{\left({\rho_{m}}+{P}\right)} }  
 + \frac{2}{3}\doublesqbraket{  \eta  {\nabla}_{\bot a} \tilde{\Theta}}\,.
 \end{eqnarray}
Putting equation \eqref{eq:junction_momentum_eqn} into equation \eqref{eq:simplified_junction_condition} and performing the angular integration gives
 \begin{eqnarray}\label{eq:full_gamma_exp}
  {\frac {d}{dt}}\bigg|_{t=0} \int_{ I}   \gamma_{a} \,{\d} S  &=&  \int_{ I}   \left( \Sigma_{+a}+ \Sigma_{{-}a} \right) 
  -\oint_{\partial I}  \doublesqbraket{\Phi  + \frac{1}{(1+w)} \frac{P}{\rho}  - \frac{2}{3} \eta \tilde{\Theta}} t_{a} \d s\,.
 \end{eqnarray}
 where we set $ P  = w \rho_{m}$ with $w=$ constant for simplicity. 
 Substituting equation \eqref{eq:full_gamma_exp} into equation \eqref{eq:vctor_cir_gamma} and performing further algebraic simplification leads to 
\begin{eqnarray}%\nonumber
{\frac {d}{dt}}\bigg|_{t=0} \Gamma_a  &=&    \int_{\partial V_{-}}\eta_{-} {n}_{b} {\D}^b \tilde{\omega}_{-a}  \d S_{-} 
%+\int_{\partial V_{-}}c_{s{-}}^2 {\Theta}_{-} \varepsilon_{ab}  {v}^b_{-} \d S_{-}
+  \int_{\partial V_{+}}\eta_{+} {n}_{b} {\D}^b \tilde{\omega}_{+a}  \d S_{+} 
%+\int_{\partial V_{+}}c_{s{+2}}^2 {\Theta}_{+} \varepsilon_{ab}  {v}^b_{+} \d S_{+}\\  &&
  -\oint_{\partial I}  \doublesqbraket{\Phi  + \frac{1}{(1+w)} \frac{P}{\rho_{m}}  - \frac{2}{3} \eta \tilde{\Theta}} t_{a} \d s\,.
  \label{eq:vctor_cir_gamma2}
  \end{eqnarray}
The circulation is generated by the relative acceleration between fluid elements on each side of the boundary. 
For $ [[{\Theta}]] = 0$,  equation \eqref{eq:vctor_cir_gamma2} reduces to 
\begin{eqnarray}%\nonumber
{\frac {d}{dt}}\bigg|_{t=0} \Gamma_a  &=&    \int_{\partial V_{-}}\eta_{-} {n}_{b} {\D}^b \tilde{\omega}_{-a}  \d S_{-} %+\int_{\partial V_{-}}c_{s{-}}^2 {\Theta}_{-} \varepsilon_{ab}  {v}^b_{-} \d S_{-}
+ \int_{\partial V_{+}}\eta_{+} {n}_{b} {\D}^b \tilde{\omega}_{+a}  \d S_{+} 
%+\int_{\partial V_{+}}c_{s{+2}}^2 {\Theta}_{+} \varepsilon_{ab}  {v}^b_{+} \d S_{+}
%\\ &&
  -\oint_{\partial I}  \doublesqbraket{\Phi  + \frac{1}{(1+w)} \frac{P}{\rho_{m}} } t_{a} \d s\,.
  \label{eq:vctor_cir_gamma3}
\end{eqnarray}
Again the difference in the integral of the gravitational potential and pressure at the boundary of the thin shell are responsible for the generation of circulation in an initially irrotational fluid. This is caused by the relative acceleration due to the viscosity between fluid elements on each side of the boundary. The viscous forces on their own do not generate circulation on the boundary, rather their role is to transfer the circulation between the boundary and the fluid interior. Finally,  the integral of the gravitational potential and pressure at the boundary generates circulation, vorticity is obtained from circulation.%using equation \eqref{eq:3Dboundary_integration}.

\section{ Implications for the standard model of cosmology}\label{sec:implications}

Although the stated aim of this project was to describe a possible mechanism for the generation of vorticity in the universe. 
Our approach to the problem builds on the knowledge from the N-body simulation where vorticity generation is linked to shell crossing singularity~\cite{Libeskind:2012ya,Jelic-Cizmek:2018gdp}.  Vorticity generation and shell-crossing singularity in cosmology are complex topics on their own, putting them together to come up with a unified picture of vorticity generation is a much more complicated task.  We have been able to put this together into a consistent picture. The key result we present here is that vorticity generation at the boundary of the causal horizon is essential for avoiding the shell-crossing singularity that appears when a given coordinate is extended beyond the causal horizon. 
In addition to this key result, we would like to discuss some obvious features and predictions that emerged from this model.

\subsection{Existence of a causal limit for massive particles}

The key feature of the universe model described here is the existence of a ``causal horizon" (boundary) for time-like geodesics. The causal horizon divides the universe into two regions: expanding and collapsing regions. Using the principle of least action we derive the equations of motion and boundary conditions applicable in each case. With these equations and boundary conditions, we showed how the vorticity is sourced by the gradients in gravitational potential and pressure across the boundary.  The vorticity flux and acoustic flux generated at the boundary then diffuse away from the boundary. 
This is a unique model of the universe that does not introduce any new free parameter, rather it makes a prediction about the existence of a causal horizon for massive particles in an expanding universe. 

In general,  causal horizons are determined by the dynamics of time-like and light-like geodesics~\cite{HawkingandEllis:1973lsss.book}. 
Within the standard model of cosmology, the well-known causal horizon is the particle horizon. It is determined by the dynamics of the light-like geodesic, it is important to note that the particle horizon indicates the maximum distance light from particles could have travelled to the observer in the age of the universe. 
 Ellis and Stoeger had earlier argued that there must exist a causal horizon in our universe that is determined by the dynamics of time-like geodesics~\cite{Ellis:2010fr}.  This causal horizon is a unique feature of general relativity.  We showed how the causal limit could be determined given a halo model for a gravitationally bound cluster. 
 Regions within the causal horizon are collapsing while regions outside are expanding. 
 Ellis and Stoeger argued that only the collapsing region contributes most significantly to the dynamics of our local environment and that the dominant interaction within this region is not mediated by massless particles, instead, they are mediated by massive particles that travel at very low speeds relative to the cosmic rest frame. The vector and tensor perturbations on an FLRW background spacetime have negligible impact on the dynamics collapsed region.
 It is the differences in speed that cause the gravitationally bound cluster to decouple from the Hubble expansion since it cannot keep up with the speed of the expanding cosmic rest frame~\cite{Ellis:2010fr, Ellis:2020kry}. 
 The dynamics of null geodesics in the presence of the time-like causal limit will be discussed elsewhere.

\subsection{Generation of of the magnetic part of the Weyl tensor}

One other unique feature of the model is the emphasise on the distinction between the frames of reference of the observer and the fluids or matter fields. 
% there cannot be an observer comoving with the fluids in both regions at the same time. 
The observer frame  is tilted with respect to the fluid frame according to 
$
\tilde{u}^a \approx  {u}^a+v^a\,,
$
where $ {u}^a$ is the observer four velocity, $\tilde{u}^a$ is the fluid four-velocity  and $v^a$ is the relative velocity between them. 
The dynamics of the fluid element in the expanding region are different from the dynamics of the fluid element in the collapsing region. The friction between the fluid elements at the boundary creates the necessary condition for the generation of vorticity at the boundary. 
 Time and ruler are determined by the observer with ${u}^a$. The vorticity we describe is associated with $\tilde{u}^a$ (see equation \eqref{eq:vorticty_transformation}). 
This is crucial because we can make use of the  Ricci identity 
$
2\nabla_{[c}\nabla_{d]} \tilde{u}_{a} = R^{e}{}_{dca} \tilde{u}_{e}
$
to relate the fluid vorticity to the existence of the magnetic part of the Weyl tensor.  The Weyl tensor is the trace-free part of the Riemann tensor, it describes the curvature of spacetime. The fluctuations in the magnetic part of the Weyl tensor have been associated with the generation of gravitational waves~\cite{Dunsby:1997fyr}.
To see how the existence of vorticity is connected to existence of the magnetic part of the Weyl tensor, we project the two free indices of the Ricci identity with $\varepsilon^{cdb}$. The further simplification of the results gives
\begin{eqnarray}\label{eq:C1}
2 \varepsilon_{acd}\tilde{\D}^{c}\tilde{\sigma}^{d}{}_{b} - \frac{4}{3} \tilde{\D}_{c}\tilde{\omega}^{c}h_{ab} + \tilde{H}_{ab} + 4 \tilde{A}_{(a}\tilde{\omega}_{b)} + 2 \tilde{\D}_{\<a}\tilde{\omega}_{b\>} = 0
\end{eqnarray}
Similarly, contracting the Ricci identity with  $\varepsilon^{cda}$  gives another constraint equation for the divergence of the vorticity vector
$
\tilde{\D}_{c}\tilde{\omega}^{c}  = \tilde{A}^b \tilde{\omega}_{b}\,.
$
In the rest frame of matter fields, $ \tilde{A}^b = 0 $, therefore  ${\D}_{c}\tilde{\omega}^{c}$ vanishes, hence
\begin{eqnarray}
2 \varepsilon_{acd}\tilde{\D}^{c}\tilde{\sigma}^{d}{}_{b} + \tilde{H}_{ab}  + 2 \tilde{\D}_{\<a}\tilde{\omega}_{b\>} = 0\,.
\end{eqnarray}
For only scalar perturbations $ \varepsilon_{acd}\tilde{\D}^{c}\tilde{\sigma}^{d}{}_{b} =0$, then equation \eqref{eq:C1} reduces to
$
 \tilde{H}_{ab}  + 2 \tilde{\D}_{\<a}\tilde{\omega}_{b\>} = 0.
$
This implies that for non-zero vorticity in the fluid, the magnetic part of the Weyl tensor is non-vanishing.  This has some consequences:
\begin{itemize}
\item Gravitational wave can be generated in the limit of the scalar perturbations if the vorticity generated at the boundary is non-zero.
% In this case, the vorticity is generated at the boundary. 
\item  The non-zero magnetic part of the Weyl tensor could have a consequence for the theory of dark matter. This was discussed in detail in ~\cite{Astesiano:2022ozl}.
\item The nonzero magnetic part of the Weyl tensor could also have implications for the generation and propagation of Maxwell's magnetic field in clusters~\cite{Ellisbook:2012}.
\end{itemize}
Finally, non-zero vorticity implies the existence of coherent helicity in the fluid interior. Helicity is usually defined as the scalar product of the fluid velocity and the vorticity vector at each point in the fluid.
The helicity measures the extent to which a flow field carries vorticity in a specific direction.

\subsection{Non-local variance in Hubble rate contributing to  acceleration in the universe }

In section \ref{sec:universe_model}, we introduced a model of the universe that treats the expanding and collapsing regions of the universe consistently with suitable junction conditions.  One unique property of this model is that both regions do not exchange mass rather the expanding regions with modes greater than the size of the collapsing region simply rescales the size of the collapsing regions~\cite{Dai:2015jaa}. 
In terms of the energy-momentum tensor,  we showed in sub-section \ref{sec:fluid_element} that the jump in the Riemann tensor manifests as viscosity in the fluid element at the boundary of the collapsing and expanding regions. In this sub-section, we argue that an observer associated will infer an accelerated expanding universe even if the rate of expansion of the expanding and collapsing regions of the universe are decelerating.  This can be seen by calculating the volume average of expansion, i.e $\tilde{\Theta}$. 
To see how this works, we start with the definition of an integral of a spacetime scalar  $S$ in a manifold with an ambient metric $g_{ab}$   \cite{Gasperini:2009mu,Heinesen:2018vjp}
\begin{eqnarray}\label{eq:integral_over_scalar}
I_{W}(S)=
\int_{\mathcal{M}_{4}}\d^4 x \sqrt{-{\rm{det}}[g_{ab}]}S(x^b)     W_{\mathcal{M}_{4}}(x^b)\,,
\end{eqnarray}
where $ W_{\mathcal{M}_{4}}$ is window function that selects slicing and foliation hypersurface and $x^b$ is the adapted coordrinate system. With respect to the model described in section \ref{sec:universe_model}, $({\mathcal{M}_{4}},g)$ corresponds to the ambient spacetime, which is a union of the manifolds ${\mathcal{M}} = {\mathcal{M}}_{-} \cup  {\mathcal{M}}_{+}$ describing the expanding and collapsing regions respectively.  A full treatment of this system will include a  consistent implementation of the boundary conditions at the level of the volume integration, however, in sub-section \ref{sec:fluid_element} we consider a limit where the effect of the boundary conditions is treated as effective fluid.  This is equivalent to decomposing  equation \eqref{eq:integral_over_scalar} into disjoint set:
\begin{eqnarray}\label{eq:integral_over_scalar_splits}
I_{W}(S)&=&I^{+}_{W}(S) +I^{-}_{W}(S)
\\
&=&\int_{\mathcal{M}_{+}}\d^4 x \sqrt{-{\rm{det}}[g^{+}_{ab}]}S(x^b)     W_{\mathcal{M}_{+}}(x^b) + \int_{\mathcal{M}_{-}}\d^4 x \sqrt{-{\rm{det}}[g^{-}_{ab}]}S(x^b)     W_{\mathcal{M}_{-}}(x^b)\,,
\end{eqnarray}
The average of  a scalar on an arbitrary manifold is  defined as 
\begin{eqnarray}
\average{S}_{W_{\mathcal{M}_{4}}}=\frac{\int_{\mathcal{M}_{4}}\d^4 x \sqrt{-{\rm{det}}[g_{ab}]}S(x^b)     W_{\mathcal{M}_{4}}(x^b)}{\int_{\mathcal{M}_{4}}\d^4 x \sqrt{-{\rm{det}}[g_{ab}]}    W_{\mathcal{M}_{4}}(x^b)} = \frac{I_{W}(S)}{\mathcal{V}}\,,
\end{eqnarray}
where ${\mathcal{V}_{\mathcal{M}_{4}}} = I_{W_{\mathcal{M}_{4}}}(1)$. With respect to equation \eqref{eq:integral_over_scalar_splits}, the average of a scalar decomposes
\begin{eqnarray}
\average{S}_{W_{\mathcal{M}_{4}}}= \frac{\mathcal{V}_{\mathcal{M}_{+}} }{\mathcal{V}_{\mathcal{M}_{4}}} 
 \average{S}_{W_{\mathcal{M}_{+}}} 
+\frac{\mathcal{V}_{\mathcal{M}_{-}} }{\mathcal{V}_{\mathcal{M}_{4}}}  \average{S}_{W_{\mathcal{M}_{-}}} \,.
\end{eqnarray}
where ${\mathcal{V}_{\mathcal{M}_{4}}}  =\mathcal{V}_{\mathcal{M}_{+}} +{ \mathcal{V}_{\mathcal{M}_{-}} } $. $ \average{S}_{W_{\mathcal{M}_{+}}}$ and $ \average{S}_{W_{\mathcal{M}_{-}}}$ are average of a scalar defined on ${\mathcal{M}_{+}}$ and ${\mathcal{M}_{-}}$ respectively. 
The window function selects a  slice and foliation 
%\begin{eqnarray}
$W_{A_{0},A,B_{0},B} = \left(V^a\nabla_{a}A\right)\delta^{(D)}\left( A_{0} - A\right) \left( B_0-B\right),$
%\end{eqnarray}
where  $ A$ defines the foliation and $B$ defines the radial extent~\cite{Anton:2023icm}.
We focus on the foliation defined by the 4-velocity of the fluid $(\tilde{u}^a)$ introduced in equation \eqref{eq:relativevelocity} as seen by a tilted observer with 4-velocity $u^a$.  The effective scale factor in the expanding and collapsing regions is defined as 
\begin{eqnarray}
a_{W_{\mathcal{M}_{+}}}(t) \equiv \left( \frac{\int \d^3 x \sqrt{{\rm{det}}[\tilde{h}^{+}](t,x^i)} W_{\Sigma_{+}}(x^i) }{  \int \d^3 x \sqrt{{\rm{det}}[\tilde{h}^{+}](t_{0},x^i)}W_{\Sigma_{+}}(x^i) }\right)\,, \qquad \qquad 
a_{W_{\mathcal{M}_{-}}}(t) \equiv \left( \frac{\int \d^3 x \sqrt{{\rm{det}}[\tilde{h}^{-}](t,x^i)} W_{\Sigma_{-}}(x^i) }{  \int \d^3 x \sqrt{{\rm{det}}[\tilde{h}^{-}](t_{0},x^i)}W_{\Sigma_{-}}(x^i) }\right)
\end{eqnarray}
where $\tilde{h}$ is the metric on the hypersurface of the fluid velocity(i.e equation \eqref{eq:hypersurface_fluid}). 
The  Hubble rate in both regions is given by
\begin{eqnarray}
H_{W_{\mathcal{M}_{+}}} \equiv \frac{1}{a_{W_{\mathcal{M}_{+}}}}\frac{ \d a_{W_{\mathcal{M}_{+}}}}{\d t}\,, \qquad \qquad 
H_{W_{\mathcal{M}_{-}}} \equiv \frac{1}{a_{W_{\mathcal{M}_{-}}}}\frac{ \d a_{W_{\mathcal{M}_{-}}}}{\d t}\,.
\end{eqnarray}
The acceleration of the ambient spacetime is then given by
\begin{eqnarray}\label{eq:non_local_variance}
\frac{1}{a_{W_{\Sigma_{3}}}}\frac{ \d^2 a_{W_{\Sigma_{3}}}}{\d t^2} = \frac{\mathcal{V}_{\Sigma_{+}} }{\mathcal{V}_{\Sigma}   } \frac{1}{a_{W_{\Sigma_{+}}}}\frac{ \d^2a_{W_{\Sigma_{+}}}}{\d t^2} +  \frac{\mathcal{V}_{\Sigma_{-}} }{\mathcal{V}_{\Sigma}   }   \frac{1}{a_{W_{\Sigma_{-}}}}\frac{ \d^2 a_{W_{\Sigma_{-}}}}{\d t^2} +   \frac{\mathcal{V}_{\Sigma_{+}} }{\mathcal{V}_{\Sigma}   } \frac{\mathcal{V}_{\Sigma_{1}} }{\mathcal{V}_{\Sigma}   }\left( H_{W_{\Sigma_{+}}} - H_{W_{\Sigma_{-}}} \right)^2\,.
\end{eqnarray}
A full treatment of this involves the use of the Einstein field equations and spatial averaging techniques~\cite{Buchert:2008zm,Umeh:2010pr,Clarkson:2011zq}, however, the point here is that the model we describe does not only describe a mechanism for the generation of vorticity in the universe, it also gives a natural explanation for accelerated expansion in the universe without the need for any assumption about the energy content of the universe. 
Here, the acceleration of the ambient spacetime volume,  ${ \d^2 a_{W_{\Sigma_{3}}}}/{\d t^2} > 0$,  is easily realised for known standard matter source, the accelerated expansion is simply due to the fact that Hubble rate in the collapsing region has a sign opposite to the Hubble rate in the expanding region $H_{W_{\Sigma_{-}}} \propto -H_{W_{\Sigma_{+}}}$.
%The last term in equation \eqref{eq:non_local_variance} is what we call the non-local variance in the Hubble rate .

\section{Conclusion}\label{sec:conc}

Our understanding of the evolution of large-scale structures in the universe is built on  an expanding FLRW background spacetime.  The general relativistic $1+3$ covariant decomposition provides tools for studying evolution of spacetimes by simply looking at how one-parameter family of geodesics propagates. One key property of geodesics in general relativity is that they can cease to be geodesics in a finite time or affine parameter~\cite{Penrose:1964wq}.  The geodesic,  the path a particle takes in a gravitational field  is determined by the curvature of spacetime, which in turn is determined by the distribution of matter and energy. The changes in the nature of matter and energy distribution are reflected on the propagation of geodesics.  
 The study of validity range of geodesics have received very little attention in cosmology even when geodesics constitute  the bedrock for studying the growth of structures using the N-body simulations~\cite{Crocce:2005xz, Springel:2005mi,Adamek:2014xba}.
Geodesics are studied more diligently in the field of differential geometry, especially in the blackholes physics where  it is well understood that presence of horizons is inevitable in general relativity~\cite{HawkingandEllis:1973lsss.book,Ashtekar:2003hk}.  
 %A geodesic could cease to be a geodesic in a finite time~\cite{Penrose:1964wq}. 
 
 In this paper, we  have studied the validity range of time-like geodesics in a universe like ours which has over-dense(gravitationally bound region, not undergoing Hubble flow) and under-dense regions (undergoing Hubble flow).  We show that geodesics on expanding region of the universe cannot be  be extended to the  over-dense region if singularities are to be avoided.
 We show that a causal horizon forms at the zero-velocity surface which then  serve as a boundary separating  families of one-parameter family of  geodesics that are causally disconnected. This distinction was pointed earlier by Ellis and Stoeger~\cite{Ellis:2010fr}.  We argue that the fact that the determinant is finite at the causal horizon allows to define consistent junction conditions for the coordinates, metrics and extrinsic curvature tensors at the boundary.  We derived a generalised  boundary condition starting from the principle of least action. The generalised  boundary condition we found reduces naturally to the Israel junctions conditions~\cite{Israel:1966rt}.

The physical picture  of what we have described mathematically can be easily be visualised as follows.  The existence of a causal horizon for time-like geodesics divides the universe into expanding and collapsing regions. The over-dense region such as the gravitationally bound systems like clusters are decoupled from the Hubble expansion. Dynamics within the gravitational bound regions are more crucial for the formation of sub-structures such as galaxies and they are causally disconnected from the dynamics in the expanding region. 
The causal horizon exists because massive particles within the over-dense region are  moving with a slower velocity such that they cannot catch up with the Hubble flow~\cite{Ellis:2010fr}.
%We described how to build a more accurate model of the observed universe that captures the dynamics in both the expanding and collapsing regions in section \ref{sec:universe_model}. 

One of the key feature of the scenario we described is that it provides a mechanism for the generation of vorticity in the universe.   The jump in the Riemann tensor at the boundary of both regions  could be interpreted in terms of the stress-energy tensor.  We show that the stress-energy tensor at the boundary  has a natural physical interpretation in terms of the effective energy-momentum tensor and therefore, can be decomposed with respect to the fundamental four vector which then allows to apply Morton's boundary layer theory for the generation of vorticity~\cite{Morton1984}.
The components of the boundary effective energy-momentum tensor include a non-vanishing  boundary anisotropic stress tensor,  pressure and the momentum flux vector. These observables exist even if the matter in the universe is purely dust. We then showed that the non-vanishing contribution of these observables leads to a non-vanishing scalar and vector circulation sourced by the jump in gradients of gravitational potential, pressure and expansion. 

The results of this study is very crucial towards understanding the local universe because the vorticity has been observed around filaments, clusters, etc~\cite{Wang:2021axr}.  Also, some high-resolution N-body simulations have been able to detect vorticity in the outskirts of the large-scale structures but the mechanism for its formation has eluded a clear analytic understanding~\cite{Libeskind:2012ya,Jelic-Cizmek:2018gdp}.  Although there are earlier works based on the N-body simulations that established a link between vorticity generation and caustics formation or shell-crossing singularity in cosmology~\cite{Pichon:1999tk,Wang:2013gtn}, This is the first paper to describe in detail how vorticity could be generated at the boundary layer between  the over-dense and under-dense regions in the universe in line with the Morton boundary layer theory~\cite{Morton1984}. 
Furthermore, we showed that the vorticity flux is sourced by the jump in gradients of the gravitational potential, pressure and rate of expansion. The vorticity flux is generated at the boundary, then gradually diffuses towards the fluid interior. 
%The presence of vorticity in the fluid especially in the collapsing region changes the structure of the Raychaudhuri equation and therefore, the focusing theorem. This indicates that the formation of caustics may be avoided provided that the fluid approach the causal limit first. 

Finally, we discussed other obvious predictions/extensions of our work in section \ref{sec:implications}. Some of these predictions/extensions include the existence of a causal horizon for massive particles in the universe.  The implications for the existence of causal horizon in relation to the possible role the magnetic part of the Weyl tensor could play in the local universe.  There is also a possibility that non-zero vorticity could act as a source for the magnetic fields in clusters. We also discussed how the non-local variance in the Hubble rate between the collapsing and expanding regions could manifest as an apparent accelerated expansion of the ambient spacetime. A comprehensive discussion of these and many more would be provided elsewhere.

\section*{Acknowledgement}
I benefited immensely from discussions with  Emir Gumrukcuoglu, Mathew Hull and  Kazuya Koyama. 
OU  is supported by the 
UK Science \& Technology Facilities Council (STFC) Consolidated Grants Grant ST/S000550/1. % \red{and ST/S000550/1}. 
The perturbation theory computations in this paper  were done with the help of tensor algebra software xLightcone and  xPand~\cite{Pitrou:2013hga},  These apps are based on xPert and xTensor~\cite{Brizuela:2008ra}.
I made use of COLOSSUS (A python toolkit for cosmology, large-scale structure, and dark matter halos) developed by Benedikt Diemer for computations involving the dark matter halo profiles~\cite{Diemer:2017bwl}.

\section*{Data Access Statement}

{No new data were generated or analysed in support of this research.}

%\bibliographystyle{$HOME/Dropbox/UWC_papers/q-dipole/draft/JHEP}%{JHEP}%
%\bibliography{$HOME/Dropbox/UWC_papers/q-dipole/draft/cosmoref.bib}
%\providecommand{\href}[2]{#2}\begingroup\raggedright

\bibliography{Separate_Jacobian_Conjecture_V10.bbl}
\end{document}